\documentclass[aps,prd,showkeys,singlecolumn,nofootinbib,floatfix]{revtex4-2}
\usepackage{amssymb}
\usepackage{amsmath}
\usepackage{graphicx}
\usepackage{dcolumn}
\usepackage{bm}
\usepackage{hyperref}
\begin{document}

\title{On the linear causality and stability of third-order relativistic dissipative fluid dynamics}
\author{C.~V.~Brito}
\email{caio\_brito@id.uff.br}
\author{G.~S.~Denicol}
\email{gsdenicol@id.uff.br}
\affiliation{Instituto de F\'{\i}sica, Universidade Federal Fluminense, UFF, Niter\'oi, 24210-346, RJ, Brazil}

\begin{abstract}
We analyze the linear causality and stability of third-order fluid dynamics considering perturbations around a global equilibrium state. We investigate the formulation derived from kinetic theory, using the Chapman-Enskog expansion, in Phys.~Rev.~C \textbf{88}, 021903 which was shown to be in excellent agreement with solutions of the microscopic theory. From this analysis, we demonstrate that this theory is linearly acausal and unstable and that such instabilities cannot be corrected by tuning the transport coefficients. We then propose a modification of this theory, valid only in the linear regime, that can be constructed to be linearly causal and stable and obtain the conditions the transport coefficients must satisfy in order for this to be the case.
\end{abstract}

\maketitle

\section{Introduction}

Relativistic dissipative fluid dynamics has been widely employed to describe the evolution of the hot and dense matter created in ultrarelativistic heavy-ion collisions \cite{gale, heinz, flork}. In these collisions a fluid is created under extreme conditions, with gradients of temperature and fluid-velocity that are not small when compared to the typical microscopic scales of the system \cite{knudsen1, knudsen2}. Naturally, the main goal of these experiments is to study the properties of the novel state of nuclear matter that is produced at the early stages of the reaction  \cite{rischke2, hatsudabook}. Nevertheless, such experiments have also triggered considerable research on how to derive relativistic formulations of fluid dynamics from microscopic theory -- a topic that remains active 
until today \cite{flork, bemfica}.    

The most intuitive approach to obtain a relativistic fluid-dynamical formulation is to extend the nonrelativistic Navier-Stokes theory, which has been successfully used to describe a wide range of fluids. However, relativistic formulations of Navier-Stokes theory \cite{eckart,landaubook} are known to be ill-defined, since they contain intrinsic linear instabilities when perturbed around an \textit{arbitrary} global equilibrium state \cite{hw0, hw2, olson, denicol2, rischke}, which renders the problem ill-posed \footnote{We remark that the authors of Ref.~\cite{liu} disagree with the conclusion that the instability and ill-posedness of relativistic Navier-Stokes theory around global equilibrium renders the theory unphysical. They argue that the solutions for the perturbations do not depend continuously on the initial data and therefore do not meet one of the three mathematical requirements to be posed with regard to realistic physical problems. So, in their reasoning, it is the type of the initial-value problem here which is physically unacceptable, and not the instability of the resulting solution.} \cite{liu}. Such linear instabilities were later shown to be related to the acausal nature of these theories \cite{hw0, denicol2, rischke}. 

Fluid-dynamical formulations that can be constructed to be linearly causal and stable were originally derived by Israel and Stewart \cite{is1, is2}, the so-called second-order theories. In contrast to Navier-Stokes theory, Israel-Stewart theory takes into account the transient dynamics of the dissipative currents, a procedure that can only be systematically achieved with the inclusion of terms which are of second order in gradients (Navier-Stokes theory only includes first-order terms). Such theories are then constructed including all possible second-order terms \cite{brsss,dnmr}, in such a way that the asymptotic solutions of the theory exactly matches, up to second order, the result expected of a gradient expansion. Israel-Stewart theory has been shown to be linearly causal and stable, as long as their transport coefficients satisfy a set of fundamental constraints \cite{hw0, olson, denicol2, rischke, bd}. More general constraints for the causality of Israel-Stewart theory, valid also in the nonlinear regime, were derived in Refs.~\cite{jn1,jn2}.

Recently, it has been argued that the inclusion of terms that are asymptotically of third order in gradients may be required to describe the dynamics of a fluid in the extreme conditions present at the early stages of heavy-ion collisions \cite{amaresh2}. For this purpose, third-order formulations of relativistic dissipative fluid dynamics were also developed \cite{el, amaresh, muronga3, amaresh3}. As a matter of fact, in Ref.~\cite{amaresh2} it was shown, assuming a Bjorken flow scenario \cite{bjorken}, that the solutions of such third-order theories are in better agreement with numerical solutions of Boltzmann equation than the solutions of Israel-Stewart theory itself when the shear viscosity is large. 

So far, the linear causality and stability of relativistic third-order fluid dynamics have not been fully explored. The goal of this work is to perform this task, analyzing linear perturbations of third-order fluid dynamics around an arbitrary global equilibrium state. For this purpose, we follow the procedure outlined in Ref.~\cite{bd}, extending it to third-order theories when required. We demonstrate that the current formulation of third-order fluid dynamics shares the same issues of relativistic Navier-Stokes theory, displaying intrinsic instabilities when perturbed around global equilibrium. We then propose a small modification to these theories that render them linearly causal and stable, and then derive conditions that constrain the values the transport coefficients can assume in order for these properties to be fulfilled. 

This paper is organized as follows: in Sec.~\ref{sec_cons} we briefly review the fundamentals of relativistic fluid dynamics. In Sec.~\ref{sec_fourier}, we consider perturbations around a global equilibrium state and linearize the fluid-dynamical equations, expressing them in Fourier space. We further decompose these equations into its transverse and longitudinal degrees of freedom, following Ref.~\cite{bd}. In Sec.~\ref{sec_third}, we analyze the linear stability of the third-order equation of motion for the shear-stress tensor proposed in Ref.~\cite{amaresh} and conclude this theory is linearly acausal and unstable for perturbations on a moving fluid. Motivated by this result, in Sec.~\ref{sec_hyper}, we then propose a modified version of this theory in order to address these problems, and obtain the set of conditions the transport coefficients must satisfy for this novel formulation to be linearly causal and stable.
All our conclusions are summarized in Sec.~\ref{sec_conc}. Throughout this paper, we use natural units, $c=k_{\mathrm{B}}=\hbar=1$, and adopt the mostly minus convention for the Minkowski metric tensor, $g^{\mu\nu}=\mathrm{diag}(+,-,-,-)$. 

\section{Conservation laws}
\label{sec_cons}

The main equations of relativistic fluid dynamics are the continuity equations related to the conservation of net-charge and energy-momentum. For the sake of simplicity, in this work we consider the net-charge to be locally zero and disregard its dynamics. Furthermore, we shall also neglect any contribution due to bulk viscous pressure. In this case, the continuity equation describing energy-momentum conservation is given by 
\begin{eqnarray}
\partial_\mu T^{\mu\nu}=\partial_\mu\left(\varepsilon u^\mu u^\nu -\Delta^{\mu\nu}P+\pi^{\mu\nu}\right)=0,
\label{eq1}
\end{eqnarray}
with $T^{\mu\nu}=\varepsilon u^\mu u^\nu -\Delta^{\mu\nu}P+\pi^{\mu\nu}$ being the energy-momentum tensor in the Landau frame \cite{landaubook}, in which the fluid 4-velocity is defined as an eigenvector of the energy-momentum tensor, $u_\mu T^{\mu\nu}=\varepsilon u^\nu$\footnote{The Landau frame corresponds to a choice of matching condition and defines the local rest frame of the fluid. In addition, one also must define the physical meaning of temperature. This is done by imposing that the energy density of the fluid in the local rest frame, $\varepsilon$, is the thermodynamic energy density and is determined in terms of the temperature by an equation of state, $\varepsilon(T)$. We remark that the choice of matching condition is arbitrary and is usually chosen by convenience.}. Above, we defined the fluid 4-velocity, $u^\mu=\gamma(1,\mathbf{V})$, with $\gamma\equiv1/\sqrt{1-V^2}$ being the Lorentz factor, the projection operator onto the 3-space orthogonal to $u^\mu$, $\Delta^{\mu\nu}=g^{\mu\nu}-u^\mu u^\nu$, the energy density, $\varepsilon$, the thermodynamic pressure, $P(\varepsilon)$, and the shear-stress tensor, $\pi^{\mu\nu}$, which is orthogonal to the fluid 4-velocity,
\begin{equation}
u^\mu \pi_{\mu\nu}=0. \label{vel_shear}
\end{equation}

It is convenient to further decompose the conservation of energy and momentum in terms of its components parallel and orthogonal to the fluid 4-velocity. In this case one obtains,
\begin{eqnarray}
u_{\nu }\partial _{\mu }T^{\mu \nu }&=&\dot{\varepsilon}+\,\left(
\varepsilon +P\right) \theta -\pi ^{\alpha \beta }\sigma _{\alpha \beta }=0,
\label{eom1} \\
\Delta _{\nu }^{\lambda }\partial _{\mu }T^{\mu \nu }&=&\left( \varepsilon
+P\right) \dot{u}^{\lambda }-\nabla ^{\lambda }P-\pi ^{\lambda \beta }\dot{u}%
_{\beta }+\Delta _{\nu }^{\lambda }\nabla _{\mu }\pi ^{\mu \nu }=0,
\label{eom2}
\end{eqnarray}
where we defined the expansion rate, $\theta\equiv\partial_\mu u^\mu$, the comoving derivative, $\dot{A}\equiv u^\mu \partial_\mu A$, the projected derivative $\nabla^\mu\equiv\Delta^{\mu\nu}\partial_\nu\equiv\partial^{\langle\mu\rangle}$, and the shear tensor as $\sigma^{\mu\nu}\equiv\Delta^{\mu\nu\alpha\beta}\partial_\alpha u_\beta$. Here, we made use of the double, symmetric, and traceless
projection operator $\Delta _{\alpha \beta }^{\mu \nu }=\left( \Delta
_{\alpha }^{\mu }\Delta _{\beta }^{\nu }+\Delta _{\beta }^{\mu }\Delta
_{\alpha }^{\nu }\right) /2-\Delta ^{\mu \nu }\Delta _{\alpha \beta }/3$. Naturally, the conservation laws provide equations of motion for the velocity field and energy density and an additional equation of motion for the shear-stress tensor is required for closure. 

\section{Linearized equations in Fourier space}
\label{sec_fourier}

We are interested in analyzing the stability of third-order fluid-dynamical formulations in their linear regime. In order to achieve this goal, we perform a \textit{linear stability analysis}, in which the system is assumed to be initially in a global equilibrium state and then we perform small perturbations on the hydrodynamic variables around such state. These perturbations can be expressed as
\begin{equation}
    \varepsilon=\varepsilon_0+\delta\varepsilon,\hspace{0.2cm}u^\mu=u^{\mu}_0+\delta u^{\mu},\hspace{0.2cm}\pi^{\mu\nu}=\delta\pi^{\mu\nu},
    \label{perturb}
\end{equation}
with $\varepsilon_0$ being the energy density of the background system and $u^{\mu}_0$ being the background 4-velocity. Since the system is taken to be initially in an equilibrium state, the shear-stress tensor is the perturbation itself. 

An essential requirement for a fluid-dynamical formulation is stability, i.e., perturbations are damped with time and thus the fluid goes back to its initial global equilibrium state some time after being perturbed. On the other hand, in an unstable theory, the perturbations would grow exponentially with time and the system will never go back to its initial equilibrium state. 

The first step here is to derive the linearized equations of motion from the conservation laws, Eqs.~(\ref{eom1}) and (\ref{eom2}), which read
\begin{eqnarray}
D_{0}\left( \frac{\delta \varepsilon }{\varepsilon _{0}+P_{0}}\right) +\nabla _{0}^{\mu
}\delta u_{\mu }& =\mathcal{O}\left( 2\right)\approx0,\label{linear_cons1} \\
D_{0}\delta u^{\mu }-\nabla _{0}^{\mu }c_s^2\left( \frac{\delta \varepsilon}{\varepsilon _{0}+P_{0}}\right)
+\nabla _{0}^{\nu }\frac{\delta\pi_{\nu}^{\mu}}{\varepsilon _{0}+P_{0}}& =\mathcal{O}\left( 2\right)\approx0,\label{linear_cons2}
\end{eqnarray}
where $D_0\equiv u_0^\mu\partial_\mu$ is the comoving derivative with respect to the background fluid velocity and $\nabla_0^\mu\equiv\Delta^{\mu\nu}_0\partial_\nu$ is the linearized projected derivative. It is possible to take the perturbations to be as small as we desire, and thus second-order or higher terms in the perturbations are neglected.

Furthermore, it is convenient to express the linearized fluid-dynamical equations in Fourier space. We adopt the following convention for the Fourier transform
\begin{eqnarray}
\tilde{M}(k^{\mu }) &=&\int d^{4}x\hspace{0.1cm}\exp \left( -ix_{\mu }k^{\mu
}\right) M(x^{\mu }), \label{transf1}\\
M(x^{\mu }) &=&\int \frac{d^{4}k}{(2\pi )^{4}}\hspace{0.1cm}\exp \left(
ix_{\mu }k^{\mu }\right) \tilde{M}(k^{\mu }),\label{transf2}
\end{eqnarray}
where $k^{\mu}=(\omega ,\mathbf{k})$, with $\omega$ being the frequency
and $\mathbf{k}$ the wave vector. As proposed in Ref.~\cite{bd}, we shall write the equations in terms of the covariant variables, 
\begin{eqnarray}
\Omega &\equiv &u_{0}^{\mu }k_{\mu }, \\
\kappa ^{\mu } &\equiv &\Delta _{0}^{\mu \nu }k_{\nu },
\end{eqnarray}
which correspond to the frequency and wave vector in the local rest frame of the unperturbed system, respectively. We also introduce the covariant wave number as
\begin{equation}
\kappa=\sqrt{-\kappa_\mu \kappa^\mu}.
\end{equation}
For the sake of convenience, throughout this work, we shall always assume that the wave vector is parallel to the unperturbed fluid velocity. Without loss of generality, we assume that the background fluid velocity is in the $x$--axis. This leads to $u^{\mu}=\gamma(1,V, 0,0)$ and $k^{\mu}=(\omega,k, 0,0)$, and hence
\begin{eqnarray}
\Omega&=&\gamma(\omega-V k), \label{omega} \\
\kappa^2&=&\gamma^2(\omega V-k)^2. \label{kappa}
\end{eqnarray}

We can now show that the Fourier transformed versions of Eqs.~(\ref{linear_cons1}) and (\ref{linear_cons2}) are written as
\begin{eqnarray}
\Omega \left(\frac{\delta\tilde{\varepsilon}}{\varepsilon _{0}+P_{0}}\right)+\kappa^{\mu}\delta \tilde{u}_{\mu}&=&0, \label{fourier1}\\
\Omega \delta \tilde{u}^{\mu}-\kappa^{\mu}c_s^2 \left(\frac{\delta\tilde{\varepsilon}}{\varepsilon _{0}+P_{0}}\right)+\kappa^{\nu}\frac{\delta\tilde{\pi}_{\nu}^{\mu}}{\varepsilon _{0}+P_{0}}&=&0. \label{fourier2} 
\end{eqnarray}
Note that these equations are valid for any fluid-dynamical formulation since the explicit form of the shear-stress tensor was not specified. Therefore, it is now necessary to introduce an equation of motion for $\pi^{\mu\nu}$ and include its linearized Fourier transformed version to Eqs.~\eqref{fourier1} and \eqref{fourier2}. Then, we can compute the dispersion relations associated with the corresponding theory and obtain the conditions that are required for such formulation to be linearly causal and stable.

\section{Third-order fluid dynamics}
\label{sec_third}

This section is dedicated to the linear stability analysis of the third-order formulation derived in Ref.~\cite{amaresh}. In this work, a third-order equation of motion for the shear-stress tensor is obtained from the relativistic Boltzmann equation using the Chapman-Enskog method, leading to the following dynamical equation for the shear-stress tensor
\begin{equation}
\dot{\pi}^{\langle\mu\nu\rangle}=2\frac{\eta}{\tau_\pi}\sigma^{\mu\nu}-\frac{1}{\tau_\pi}\pi^{\mu\nu}+\frac{4}{35}\nabla^{\langle\mu}\left(\tau_\pi\nabla_\alpha \pi^{\nu\rangle\alpha}\right)-\frac{2}{7}\nabla_{\alpha}\left(\tau_\pi \nabla^{\langle\mu}\pi^{\nu\rangle\alpha}\right)-\frac{1}{7}\nabla_\alpha\left(\tau_\pi\nabla^{\alpha}\pi^{\langle\mu\nu\rangle}\right)+\cdots, \label{eq_amaresh_dots}
\end{equation}
which corresponds to Eq.~(16) of the aforementioned paper. Note that the dots represent terms that are nonlinear and thus do not contribute to the linear stability analysis that follows and, hence, were omitted for the sake of convenience. Naturally, a third-order equation of motion for the shear-stress tensor has several additional terms that are not present in the Israel-Stewart equations. In particular, the last three terms on the right-hand side of Eq.~\eqref{eq_amaresh_dots} are linear contributions of third-order which are absent in Israel-Stewart theory. The primary goal of this section is to analyze the effects that the inclusion of these terms have in the linear causality and stability of the theory, since these properties are well known when such terms are not present \cite{hw0, olson, rischke, bd}.

Considering the perturbations introduced in Eq.\ \eqref{perturb}, the linearized third-order equation of motion for the shear-stress tensor becomes
\begin{equation}
D_0 \delta\pi^{\mu\nu}+\frac{1}{\tau_\pi}\delta\pi^{\mu\nu}=\frac{\eta}{\tau_\pi}\left[2\nabla_0^{(\mu}\delta u^{\nu)}-\frac{2}{3}\Delta_0^{\mu\nu}\partial_\lambda \delta u^\lambda\right]-\frac{1}{7}\tau_\pi\nabla^0_{\lambda}\nabla_0^{\lambda}\delta\pi^{\mu\nu}-\frac{6}{35}\tau_\pi\nabla^0_\lambda\left[2\nabla_0^{(\mu}\delta\pi^{\nu)\lambda}-\frac{1}{3}\Delta_0^{\mu\nu}\nabla^0_\beta\delta\pi^{\beta\lambda}\right].\label{shear_third}
\end{equation}
The last two terms on the right-hand side of the equation above are corrections due to third-order contributions in the equation of motion for the shear-stress tensor. At this point, one may ask whether the inclusion of third-order terms in the equation for the shear-stress tensor either maintains the properties of linear causality and stability observed in Israel-Stewart theory or yields novel constraints for the transport coefficients. 

In order to study the modes of the theory, it is first necessary to calculate the linearized third-order equation of motion for the shear-stress tensor in Fourier space, which reads
\begin{eqnarray}
i\Omega\delta\tilde{\pi}^{\mu\nu}+\frac{1}{\tau_\pi}\delta\tilde{\pi}^{\mu\nu}&=&i\frac{\eta}{\tau_\pi}\left[2\kappa^{(\mu}\delta\tilde{u}^{\nu)}-\frac{2}{3}\Delta_0^{\mu\nu}\kappa_\lambda \delta\tilde{u}^{\lambda}\right]-\frac{1}{7}\tau_\pi \kappa^2 \delta\tilde{\pi}^{\mu\nu}+\frac{6}{35}\kappa_\lambda\tau_\pi\left[\kappa^{(\mu}\delta\tilde{\pi}^{\nu)\lambda}-\frac{1}{3}\Delta_0^{\mu\nu}\kappa_\beta \delta\tilde{\pi}^{\beta\lambda}\right].\label{shear_four}
\end{eqnarray}

The linear stability analyses performed throughout this work will follow the procedure originally developed in Ref.~\cite{bd}, in which the linearized fluid-dynamical equations in Fourier space are decomposed into their transverse (orthogonal to $\kappa^\mu$) and longitudinal (parallel to $\kappa^\mu$) components. 
Such components decouple in the linear regime, and the equations associated to each of them can be solved independently, considerably simplifying the calculations. For this purpose, we define the following projection operator in Fourier onto the 3-space orthogonal to $\kappa^{\mu}$,
\begin{equation}
\Delta^{\mu\nu}_\kappa\equiv g^{\mu\nu}+\frac{\kappa^\mu \kappa^\nu}{\kappa^2}.\label{proj_fourier}
\end{equation}
Then, a 4-vector can be decomposed in its transverse and longitudinal components with respect to the wave 4-vector in the local rest frame of the background system as
\begin{equation}
A^\mu=A_\|\kappa^\mu+A^\mu_\bot,\label{rank1_fourier}
\end{equation}
with the longitudinal component being defined as $A_\|=-\kappa_\mu A^\mu/\kappa$ while the transverse component is $A^\mu_\bot=\Delta^{\mu\nu}_{\kappa}A_\nu$. A similar approach can be performed to decompose a traceless rank two tensor. In this case, it is first essential to introduce the double, symmetric, and traceless projection operator in Fourier space as
\begin{equation}
\Delta^{\mu\nu\alpha\beta}_{\kappa}=\frac{1}{2}\left(\Delta^{\mu\alpha}_{\kappa}\Delta^{\nu\beta}_{\kappa}+\Delta^{\mu\beta}_{\kappa}\Delta^{\nu\alpha}_{\kappa}\right)-\frac{1}{3}\Delta^{\mu\nu}_\kappa\Delta^{\alpha\beta}_{\kappa}.\label{dproj_fourier}
\end{equation}
Wherefore, the decomposition of an arbitrary traceless rank two tensor $A^{\mu\nu}$ can be performed as follows
\begin{equation}
A^{\mu\nu}=A_{\|}\frac{\kappa^{\mu}\kappa^{\nu}}{\kappa^{2}}+\frac{1}{3}A_{\|}\Delta_{\kappa}^{\mu\nu}+A_{\bot}^{\mu}\frac{\kappa^{\nu}}{\kappa}+A_{\bot}^{\nu}\frac{\kappa^{\mu}}{\kappa}+A_{\bot}^{\mu\nu},\label{rank2_fourier}
\end{equation}
with its respective projections being defined as $A_{\|}\equiv\kappa_{\mu}\kappa_{\nu}A^{\mu\nu}/\kappa^{2}$, $A_{\bot}^{\mu}\equiv-\kappa^{\lambda}\Delta_{\kappa}^{\mu\nu}A_{\lambda\nu}/\kappa$, and $A_{\bot}^{\mu\nu}\equiv\Delta_{\kappa}^{\mu\nu\alpha\beta}A_{\alpha\beta}$.
Thereby, the linearized equations for the conservation laws in Fourier space, Eqs.~(\ref{fourier1}) and (\ref{fourier2}), and the linearized equation of motion for the shear-stress tensor in Fourier space, Eq.~(\ref{shear_four}), can be split into two different components that decouple and can be solved independently, as it will be shown in the next sections. It is then possible to obtain the dispersion relations associated with the transverse and longitudinal modes.

\subsection{Transverse modes}

In this section, we shall obtain and analyze the relevant transverse modes of the third-order theory. First, note that Eq.~(\ref{fourier1}) is a scalar equation and therefore it has no transverse component. On the other hand, the transverse component of Eqs.~(\ref{fourier2}) is obtained projecting it with $\Delta_{\mu,\kappa}^{\lambda}$, and it is given by
\begin{equation}
\hat{\Omega}\delta\tilde{u}^\lambda_\bot-\hat{\kappa}\frac{\delta\tilde{\pi}^\lambda_{\bot}}{\varepsilon_0+P_0}=0.  \label{eom_trans1}
\end{equation}
For the sake of convenience, the variables here are rescaled in terms of the hydrodynamic timescale $\tau_\eta\equiv\eta/(\varepsilon_0+P_0)$ in order to work only with dimensionless variables, $\hat{A}=A[\tau_\eta]$. Note that this equation couples with the partially transverse component of $\delta\tilde\pi^{\mu\nu}$. We obtain the equation of motion for this component by projecting Eq.~(\ref{shear_four}) with $\kappa_\mu\Delta^{\lambda}_{\nu,\kappa}$, leading to
\begin{equation}
\left(i\hat{\tau}_{\pi}\hat{\Omega}+\frac{8}{35}\hat{\tau}_{\pi}^2\hat{\kappa}^2+1\right)\frac{\delta\tilde{\pi}^\lambda_{\bot}}{\varepsilon_0+P_0}-i\hat{\kappa}\delta\tilde{u}^\lambda_\bot=0.
\end{equation}
These equations can be written in the following matrix form
\begin{equation}
\left( 
\begin{array}{cc}
i\hat{\tau}_{\pi}\hat{\Omega}+\frac{8}{35}\hat{\tau}_{\pi}^2\hat{\kappa}^2+1 & -i\hat{\kappa}  \\ 
-\hat{\kappa} & \hat{\Omega} 
\end{array}%
\right) \left( 
\begin{array}{c}
\frac{\delta\tilde{\pi}^\mu_{\bot}}{\varepsilon_0+P_0} \\
\delta \hat{u}_{\bot}^{\mu }
\end{array}%
\right)=0,
\end{equation}
which leads to the following dispersion relation
\begin{equation}
\hat{\Omega}\left(i\hat{\tau}_{\pi}\hat{\Omega}+\frac{8}{35}\hat{\tau}_{\pi}^2\hat{\kappa}^2+1\right)-i\hat{\kappa}^2=0.\label{disp_trans}
\end{equation}
Note that if the quadratic term inside the parentheses which accounts for all linear third-order contributions is set to zero, one immediately recovers the dispersion relation satisfied by the transverse modes of Israel-Stewart theory considering only dissipation via shear-stress, see Refs.~\cite{rischke, bd}. 

The next step is to analyze whether the modes from the third-order formulation proposed in Ref.~\cite{amaresh} are linearly stable and if the occurrence of additional terms in the equation of motion for the shear-stress tensor yields new constraints for the transport coefficients. 

\subsubsection{Perturbations on a static fluid}

The first case that will be studied here is the one for perturbations performed in a fluid at rest, i.e., when $V=0$ in Eqs.~(\ref{omega}) and (\ref{kappa}). In this case, $\Omega=\omega$ and $\kappa=k$, and the dispersion relation associated with the transverse modes becomes simply
\begin{equation}
\hat{\omega}\left(i\hat{\tau}_{\pi}\hat{\omega}+\frac{8}{35}\hat{\tau}_{\pi}^2\hat{k}^2+1\right)-i\hat{k}^2=0.\label{static_trans}
\end{equation}

\begin{figure}[ht]
\begin{center}
\includegraphics[width=.9\textwidth]{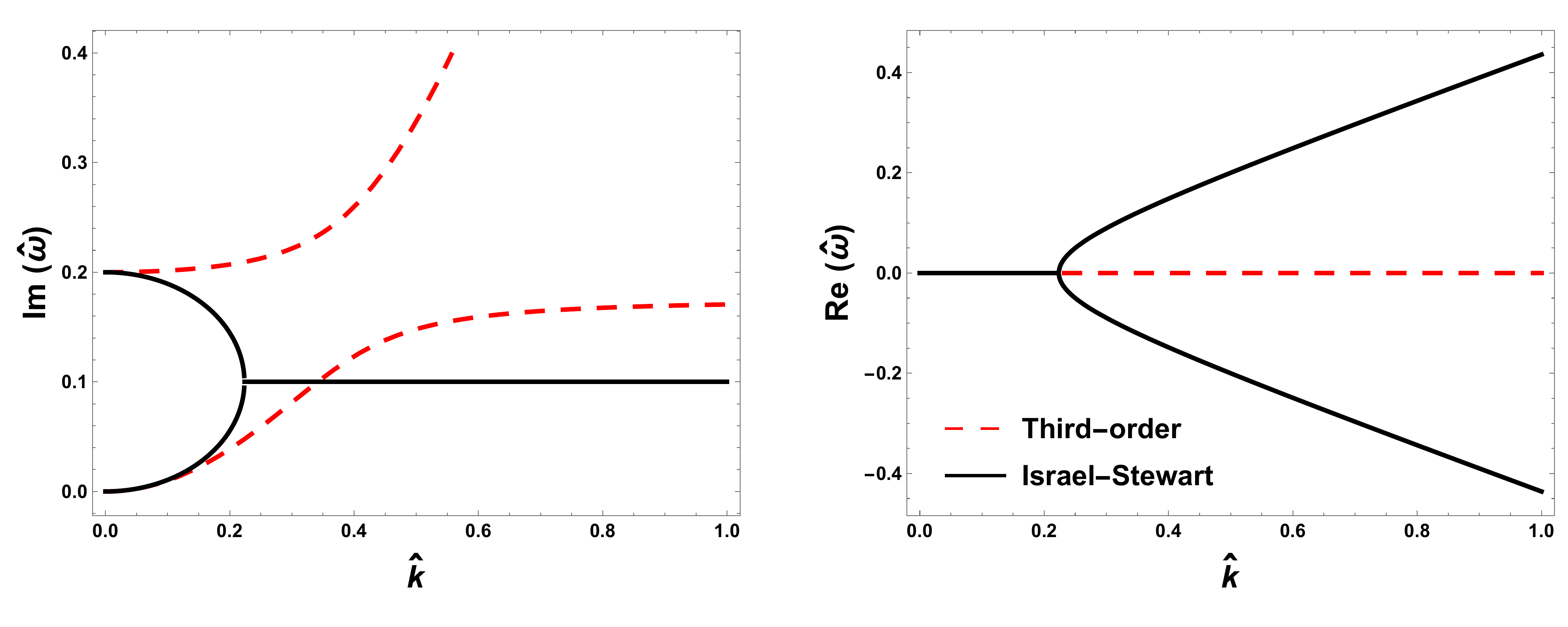}
\caption{Imaginary and real parts of the transverse modes of Israel-Stewart theory (solid black lines) and third-order fluid dynamics (red dashed lines), considering $\hat{\tau}_\pi=5$ \cite{dnmr}, for perturbations on a static background fluid.}
\label{is_amaresh_strans}
\end{center}
\end{figure}

The solutions of this equation are
\begin{eqnarray}
\hat{\omega}_{T,\pm}^{\mathrm{shear}}=i\left[\frac{1+\frac{8}{35}\hat{\tau}_{\pi}^2\hat{k}^2\pm\sqrt{\left(1+\frac{8}{35}\hat{\tau}_{\pi}^2\hat{k}^2\right)^2-4\hat k^2 \hat\tau_\pi}}{2\hat\tau_\pi}\right].
\end{eqnarray}
In the small wave number limit, the transverse modes of the theory can be written as
\begin{eqnarray}
\hat\omega^{\mathrm{shear}}_{T,-}&=&i\hat k^2+i\hat{\tau}_{\pi}\left(1-\frac{8\hat\tau_\pi}{35}\right)\hat k^4+\mathcal{O}(\hat k^6),\label{trans_stat1} \\
\hat\omega^{\mathrm{shear}}_{T,+}&=&\frac{i}{\hat\tau_\pi}-i\left(1-\frac{8\hat\tau_\pi}{35}\right)\hat k^2+\mathcal{O}(\hat k^4). \label{trans_stat2}
\end{eqnarray}
Thus, there is the occurrence of one hydrodynamic and one nonhydrodynamic mode and, in the small $\hat{k}$ limit, they are identical to the transverse modes of Israel-Stewart theory \cite{bd}. In the large wave number limit, we have the following asymptotic expansion,  
\begin{eqnarray}
\hat\omega^{\mathrm{shear}}_{T,-}&=&\frac{35}{8\hat\tau_\pi^2}i-\left(\frac{35}{8 \hat\tau_\pi^2}\right)^2\left(1-\frac{35}{8\hat\tau_\pi}\right)\frac{i}{\hat k^2} +\mathcal{O}\left(\frac{1}{\hat k^4}\right),\\
\hat\omega^{\mathrm{shear}}_{T,+}&=&\frac{8}{35}i\hat\tau_\pi \hat k^2+\frac{i}{8\hat\tau_\pi}\left(1-\frac{35}{8\hat\tau_\pi}\right)+\mathcal{O}\left(\frac{1}{\hat k^2}\right).
\end{eqnarray}
We thus see that the hydrodynamic mode, $\hat\omega^{\mathrm{shear}}_{T,-}$, becomes constant at asymptotically large values of wave number, a behavior also observed in Israel-Stewart theory. However, unlike the shear modes of Israel-Stewart theory, the nonhydrodynamic mode of the third-order theory proposed in Ref.~\cite{amaresh} does not saturate at large wave number, i.e., it keeps increasing as the wave number increases. As a matter of fact, the nonhydrodynamic mode, $\hat\omega^{\mathrm{shear}}_{T,+}$, behaves as $\omega\sim ik^2$, as $k \rightarrow \infty$ -- a behavior often associated with parabolic theories. 
At this point, we can at least conclude that the theory is not hyperbolic and, thus, should display an acausal behavior. These modes are displayed in Fig.~\ref{is_amaresh_strans} (red dashed lines), considering $\hat\tau_\pi=5$ \cite{dnmr}, where we also display the shear modes of Israel-Stewart theory (solid black lines), for the sake of comparison. We finally note that, for perturbations on a static fluid, the transverse modes of the third-order theory are purely imaginary and stable. 

\subsubsection{Perturbations on a moving fluid}

In order to analyze a more interesting, and intrinsically relativistic, case, we now consider perturbations on a \textit{moving} fluid. This is the case in which instabilities usually arise in relativistic Navier-Stokes theory \cite{hw0, hw2} and it is important to verify whether they will appear in the third-order theory presented here as well. As it was already mentioned, we assume that the background fluid velocity is parallel to the wave vector. In this case, the dispersion relation that must be solved is obtained by inserting Eqs.~(\ref{omega}) and (\ref{kappa}) into Eq.~(\ref{disp_trans}), leading to 
\begin{equation}
\gamma(\hat{\omega}-V\hat k)\left[i\hat{\tau}_{\pi}\gamma(\hat{\omega}-V\hat k)+\frac{8}{35}\hat{\tau}_{\pi}^2\gamma^2(\hat\omega V-\hat k)^2+1\right]-i\gamma^2(\hat\omega V-\hat k)^2=0. \label{new_eq}
\end{equation}
One can immediately note that perturbations on top of a moving fluid lead to the occurrence of an additional mode. This is a remarkably problematic feature usually carried by parabolic formulations and is also observed in relativistic Navier-Stokes theory and the relativistic diffusion equation.

The general solutions for these modes can in principle be found, but are rather complicated in form. For the sake of convenience, we initially restrict our discussion to the asymptotic form of these modes. In the small wave number limit, we obtain the following three solutions,
\begin{eqnarray}
\hat\omega^{\mathrm{shear}}_{T,-}&=&V\hat k+\frac{i\hat k^2}{\gamma}+\mathcal{O}(\hat k^3), \label{trans_mov1}\\
\hat\omega^{\mathrm{shear}}_{T,+}&=&\frac{35i}{16\gamma\hat\tau_\pi^2 V^2}\left[(V^2-\hat\tau_\pi)+\sqrt{(V^2-\hat\tau_\pi)^2+\frac{32}{35}V^2\hat\tau_\pi^2}\right] + \mathcal{O}(\hat k), \label{trans_mov2}\\
\hat\omega^{\mathrm{shear}}_{T,\mathrm{new}}&=&\frac{35i}{16\gamma\hat\tau_\pi^2 V^2}\left[(V^2-\hat\tau_\pi)-\sqrt{(V^2-\hat\tau_\pi)^2+\frac{32}{35}V^2\hat\tau_\pi^2}\right] + \mathcal{O}(\hat k). \label{trans_mov3}
\end{eqnarray}
The first two modes correspond to boosted versions of the modes found in \eqref{trans_stat1} and \eqref{trans_stat2}. This can be immediately seen by taking the limit $V\rightarrow0$ in the equations above, which effectively recovers solutions \eqref{trans_stat1} and \eqref{trans_stat2}. The mode $\omega^{\mathrm{shear}}_{T,\mathrm{new}}$ has no analogue for perturbations around a fluid at rest and, in this sense, is novel. Linear stability dictates that the imaginary part of these modes must be positive for all possible physical values of the background velocity, especially in the vanishing wave number limit. Otherwise, homogeneous perturbations in a given system would increase exponentially, a clearly nonphysical behavior. Since the term inside the square root in the above equation is positive definite and greater than the term outside it, this new mode has always a negative imaginary part at $\hat k =0$ and, thus, is unstable. 

\begin{figure}[ht]
\begin{center}
\includegraphics[width=\textwidth]{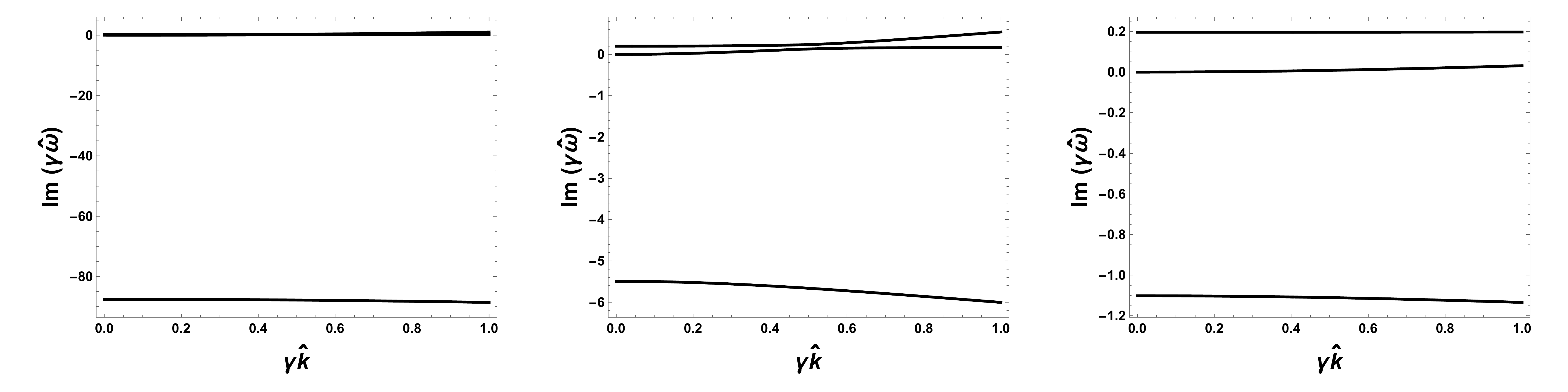}\\
\includegraphics[width=\textwidth]{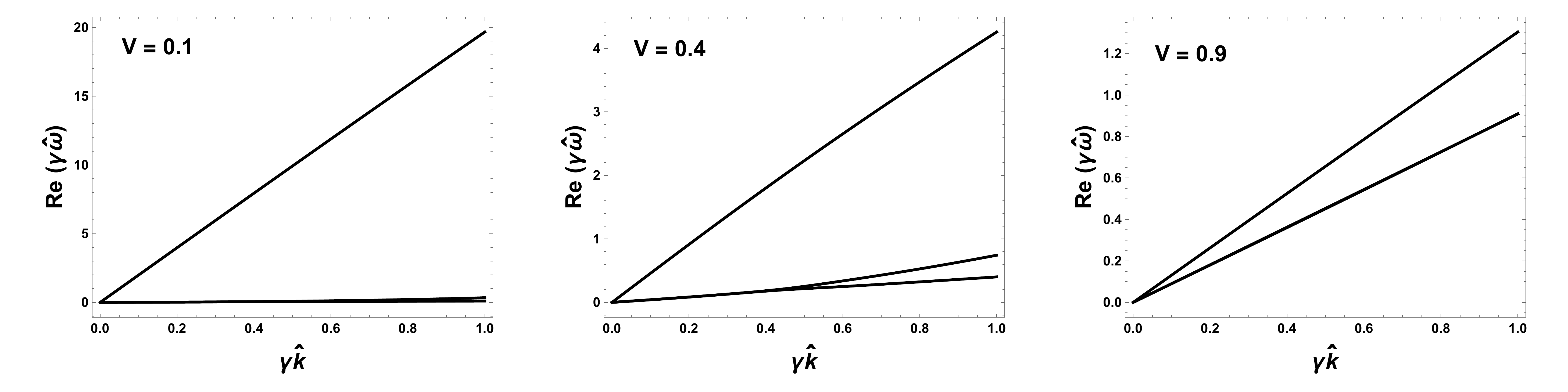}
\caption{Imaginary and real parts of the transverse modes, considering $\hat{\tau}_\pi=5$ \cite{dnmr}, for three different values of background velocity $V=0.1$, $V=0.4$, and $V=0.9$.}
\label{parab_nonstatic_trans}
\end{center}
\end{figure}

For the sake of illustration, we plot the transverse modes for perturbations on a moving fluid, solutions of Eq.~\eqref{new_eq}, as a function of the wave number $k$ in Fig.~\ref{parab_nonstatic_trans}, considering $\hat\tau_\pi=5$ \cite{dnmr}, for several values of background velocity. We see that the additional mode that appears for perturbations on a moving fluid remains unstable not only for vanishing wave number, but also for a wide range of $k$.

The acausality and instability observed on the new nonhydrodynamic mode for perturbations on a moving fluid, $\hat\omega^{\mathrm{shear}}_{T,\mathrm{new}}$, occur independently of the choice of transport coefficients and thus cannot be fixed by simply adjusting their values.

\subsection{Longitudinal modes}

We now study the dispersion relations associated with the longitudinal modes of the theory. The longitudinal projections of Eqs.~(\ref{fourier2}) and (\ref{shear_four}) are obtained by projecting them with $\kappa_\mu$ and $\kappa_\mu \kappa_\nu$, respectively. Note that Eq.~(\ref{fourier1}) is already written in terms of longitudinal components. Hence, the equations of motion related to the longitudinal degrees of freedom can be summarized as follows
\begin{eqnarray}
\hat{\Omega}\frac{\delta\tilde{\varepsilon}}{\varepsilon _{0}+P_{0}}-\hat{\kappa}\delta\tilde{u}_{\|}&=&0, \label{eom_long1}\\
\hat{\Omega}\delta\tilde{u}_{\|}-\hat{\kappa}c_s^2\frac{\delta\tilde{\varepsilon}}{\varepsilon _{0}+P_{0}}-\hat{\kappa}\frac{\delta\tilde{\pi}_{\|}}{\varepsilon_0+P_0}&=&0,\label{eom_long2}\\
\left(i\hat{\tau}_{\pi}\hat{\Omega}+\frac{9}{35}\hat{\tau}_{\pi}^2\hat{\kappa}^2+1\right)\frac{\delta\tilde{\pi}_{\|}}{\varepsilon_0+P_0}-\frac{4i}{3}\hat{\kappa}\delta\tilde{u}_\|&=&0.
\end{eqnarray}
It is possible to write the equations for the longitudinal modes in the following matrix form
\begin{equation}
\left( 
\begin{array}{ccc}
i\hat{\tau}_{\pi}\hat{\Omega}+\frac{9}{35}\hat{\tau}_{\pi}^2\hat{\kappa}^2+1 & -i\frac{4}{3}\hat{\kappa} & 0  \\ 
-\hat{\kappa} & \hat{\Omega} & -c_{\mathrm{s}}^2 \hat{\kappa} \\
0 & -\hat{\kappa} & \hat{\Omega}
\end{array}%
\right) \left( 
\begin{array}{c}
\frac{\delta\tilde{\pi}_{\|}}{\varepsilon_0+P_0} \\ 
\delta \tilde{u}_{\|} \\
\frac{\delta\varepsilon}{\varepsilon_0+P_0}
\end{array}%
\right)=0,
\end{equation}
where nontrivial solutions are obtained when the determinant vanishes, leading to the following dispersion relation,
\begin{equation}
\left(\hat{\Omega}^2-c_s^2\hat{\kappa}^2\right)\left(i\hat{\tau}_{\pi}\hat{\Omega}+\frac{9}{35}\hat{\tau}_{\pi}^2\hat{\kappa}^2+1\right)-\frac{4i}{3}\hat{\Omega}\hat{\kappa}^2=0.\label{disp_long}
\end{equation}
Note that if the term that contains third-order contributions is set to zero, one immediately recovers the dispersion relation of Israel-Stewart theory when only dissipation via shear-stress tensor is considered \cite{rischke, bd}. We shall now analyze in detail the solutions of this equation.

\subsubsection{Perturbations on a static fluid}

As it was done so far in this work, it is convenient to begin the linear stability analysis by looking at the modes for perturbations on a static background fluid. In this case, the dispersion relation associated with the longitudinal modes, Eq.~(\ref{disp_long}), then reads
\begin{equation}
\left(\hat{\omega}^2-c_s^2\hat{k}^2\right)\left(i\hat{\tau}_{\pi}\hat{\omega}+\frac{9}{35}\hat{\tau}_{\pi}^2\hat{k}^2+1\right)-\frac{4i}{3}\hat{\omega}\hat{k}^2=0. \label{static_long}
\end{equation}
In the small wave number, the longitudinal modes of the theory can be expressed as
\begin{eqnarray}
\hat\omega^{\mathrm{sound}}_\pm&=&\pm c_{\mathrm{s}}\hat k+\frac{2i}{3}\hat k^2\pm\frac{2\left(3 \hat\tau_\pi c_{\mathrm{s}}^2-1\right)}{9c_{\mathrm{s}}}\hat k^3+\mathcal{O}(\hat k^4), \label{parabolic_long_stat1}\\
\hat\omega^{\mathrm{shear}}_L&=&\frac{i}{\hat\tau_\pi}+i\left(\frac{9\hat\tau_\pi}{35}-\frac{4}{3}\right)\hat k^2+\mathcal{O}(\hat k^4). \label{shear_L_para}
\end{eqnarray}
Note the occurrence of two hydrodynamic and one nonhydrodynamic mode, which are identical to the longitudinal modes of Israel-Stewart theory \cite{bd} in the small wave number limit. Furthermore, in the large wave number limit the longitudinal modes of the third-order theory read
\begin{eqnarray}
\hat\omega^{\mathrm{sound}}_\pm&=&\pm c_{\mathrm{s}}\hat k+\frac{70i}{27\hat\tau_\pi^2}\pm\frac{2450\left(3 \hat\tau_\pi c_{\mathrm{s}}^2-1\right)}{729\hat\tau_\pi^4 c_{\mathrm{s}}\hat k},\\
\hat\omega^{\mathrm{shear}}_L&=&\frac{9i}{35}\hat\tau_\pi \hat k^2+\frac{i\left(27\hat\tau_\pi-140\right)}{27\hat\tau_\pi^2}+\mathcal{O}\left(\frac{1}{\hat k^2}\right).
\end{eqnarray}
Surprisingly, due to the inclusion of third-order terms, in the large wave number limit, the dominant part of the hydrodynamic modes, $\hat \omega^{\mathrm{sound}}_\pm$, is also $\pm c_{\mathrm{s}} \hat k$ -- thus, the asymptotic group velocity associated with these modes is always subluminal. Also, as it was first observed for the transverse modes of the theory, we see that the nonhydrodynamic mode behaves as $\hat\omega^{\mathrm{shear}}_L\sim i\hat k^2$ in the large wave number limit and thus keeps increasing as the wave number is increased, a behavior commonly displayed by parabolic formulations, such as the diffusion equation.

\begin{figure}[ht]
\begin{center}
\includegraphics[width=0.9\textwidth]{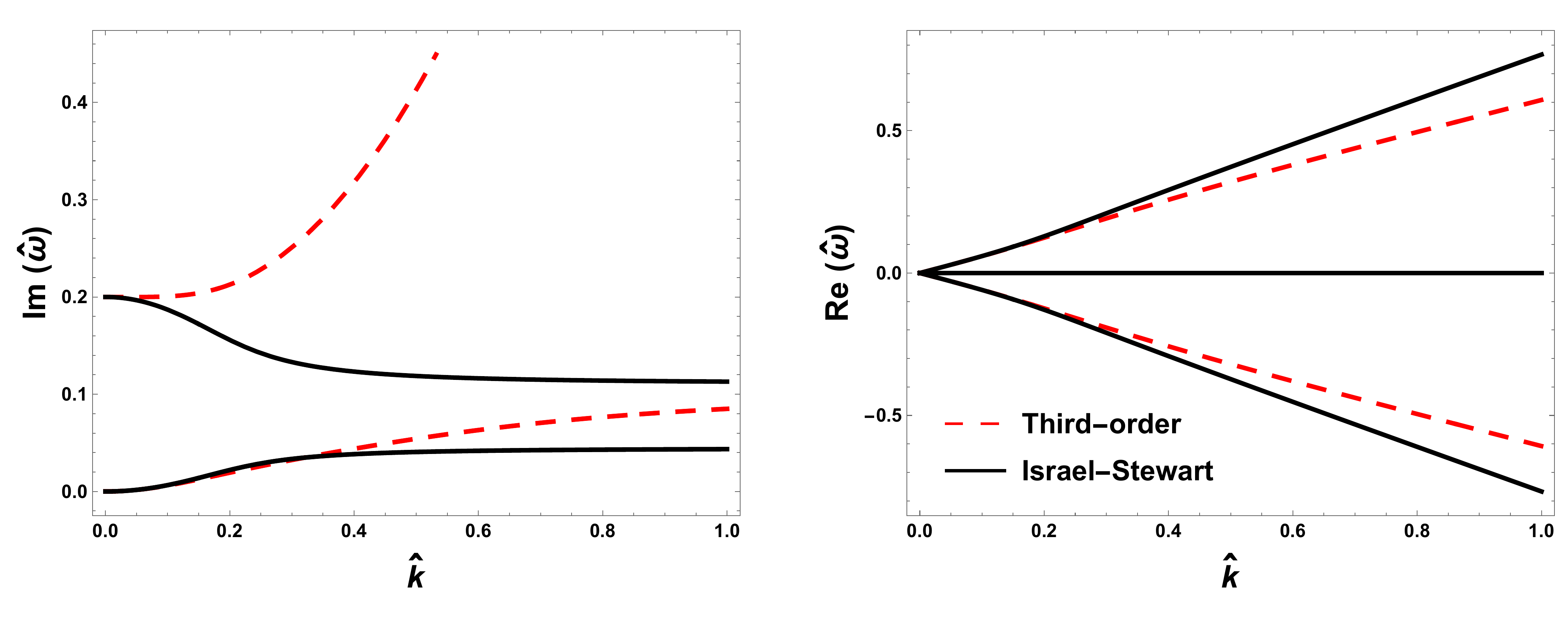}
\caption{Imaginary and real parts of the longitudinal modes of Israel-Stewart theory (solid black lines) and third-order fluid dynamics (red dashed lines), considering $\hat{\tau}_\pi=5$ \cite{dnmr}, for perturbations on a static background, in the ultrarelativistic regime, $c_s^2=1/3$.}
\label{is_amaresh_long}
\end{center}
\end{figure}

The solutions of Eq.~\eqref{static_long} are displayed in Fig.~\ref{is_amaresh_long} (red dashed lines), where we also display the longitudinal modes of Israel-Stewart theory (black solid lines), considering $\hat\tau_\pi=5$, for the sake of a quantitative comparison. There are two hydrodynamic modes, which have degenerated imaginary parts, and a single nonhydrodynamic mode, as it can be seen in the left panel. As it can be observed from Figs.~\ref{is_amaresh_strans} and \ref{is_amaresh_long}, it is possible to conclude that both transverse and longitudinal modes of the third-order formulation discussed here, respectively, are well behaved for perturbations on a static background fluid.

\subsubsection{Perturbations on a moving fluid}

We now consider perturbations on a moving fluid. As it was done so far in this work, we consider a background fluid velocity that is parallel to the wave vector, which further leads to Eqs.~(\ref{omega}) and (\ref{kappa}). In this case, the dispersion relation associated with the longitudinal modes reads
\begin{equation}
\left[\gamma^2(\hat{\omega}-V \hat k)^2-c_s^2\gamma^2(\hat\omega V-\hat k)^2\right]\left[i\hat{\tau}_{\pi}\gamma(\hat{\omega}-V \hat k)+\frac{9}{35}\hat{\tau}_{\pi}^2\gamma^2(\hat\omega V-\hat k)^2+1\right]-\frac{4i}{3}\gamma^3(\hat{\omega}-V \hat k)(\hat\omega V-\hat k)^2=0. \label{non-static_long}
\end{equation}
As it was also observed for the transverse modes, perturbations on top of a moving fluid yield an additional nonhydrodynamic mode. Thus, there are now two hydrodynamic and two nonhydrodynamic modes. Obtaining these modes for arbitrary values of wave number can be an extremely complicated task, and we shall retain ourselves to the study of their asymptotic limits, as it was done in the analysis of the transverse modes of the theory. In the small wave number limit, the longitudinal modes read
\begin{eqnarray}
\hat\omega^{\mathrm{sound}}_\pm&=&\frac{V\pm c_{\mathrm{s}}}{1\pm c_\mathrm{s}V}\hat k+\mathcal{O}(\hat k^2), \label{long_rel_vel}\\
\hat\omega^{\mathrm{shear}}_{L}&=&\frac{35i}{54\gamma}\left[\frac{3\hat{\tau}_{\pi}-4V^2-3\hat{\tau}_{\pi}V^2c_s^2-\sqrt{(3\hat{\tau}_{\pi}-4V^2-3\hat{\tau}_{\pi}V^2c_s^2)^2+\frac{324}{35}(1-c_s^2 V^2)^2 \hat{\tau}_{\pi}^2 V^2}}{\hat{\tau}_{\pi}^2 V^2 (c_s^2 V^2-1)}\right]+\mathcal{O}(\hat k),\\
\hat\omega^{\mathrm{shear}}_{L,\mathrm{new}}&=&\frac{35i}{54\gamma}\left[\frac{3\hat{\tau}_{\pi}-4V^2-3\hat{\tau}_{\pi}V^2c_s^2+\sqrt{(3\hat{\tau}_{\pi}-4V^2-3\hat{\tau}_{\pi}V^2c_s^2)^2+\frac{324}{35}(1-c_s^2 V^2)^2 \hat{\tau}_{\pi}^2 V^2}}{\hat{\tau}_{\pi}^2 V^2 (c_s^2 V^2-1)}\right]+\mathcal{O}(\hat k).
\end{eqnarray}

The hydrodynamic modes for perturbations on a moving fluid can be understood using the relativistic velocity addition rule. On the other hand, the nonhydrodynamic modes are less trivial, since we even see the appearance of a new solution. As already stated, linear stability dictates that the imaginary part of these modes must be positive for all possible values the background velocity can assume, especially in the vanishing wave number limit. Therefore, the stability of these modes is guaranteed if both numerator and denominator have the same sign. We then require that neither the numerator nor the denominator change their signs for any value of the background velocity in the causal interval, $0\leq V\leq1$, otherwise leading to a problematic discontinuity in the modes. In this case, it is straightforward to see that the denominator must be always negative. Therefore, a stable mode is obtained if the numerator is negative as well. However, since the term inside the square root is always greater than the term outside it, the numerator is positive definite and thus the mode $\hat\omega^{\mathrm{shear}}_{L,\mathrm{new}}$ is always unstable. This mode is exactly the new solution that appears when considering perturbations on a moving fluid. As it was first observed for the transverse mode $\hat\omega^{\mathrm{shear}}_{T,\mathrm{new}}$, the instability of the mode $\hat\omega^{\mathrm{shear}}_{L,\mathrm{new}}$ cannot be fixed by tuning any of the transport coefficients existing in the theory.

\begin{figure}[ht]
\begin{center}
\includegraphics[width=\textwidth]{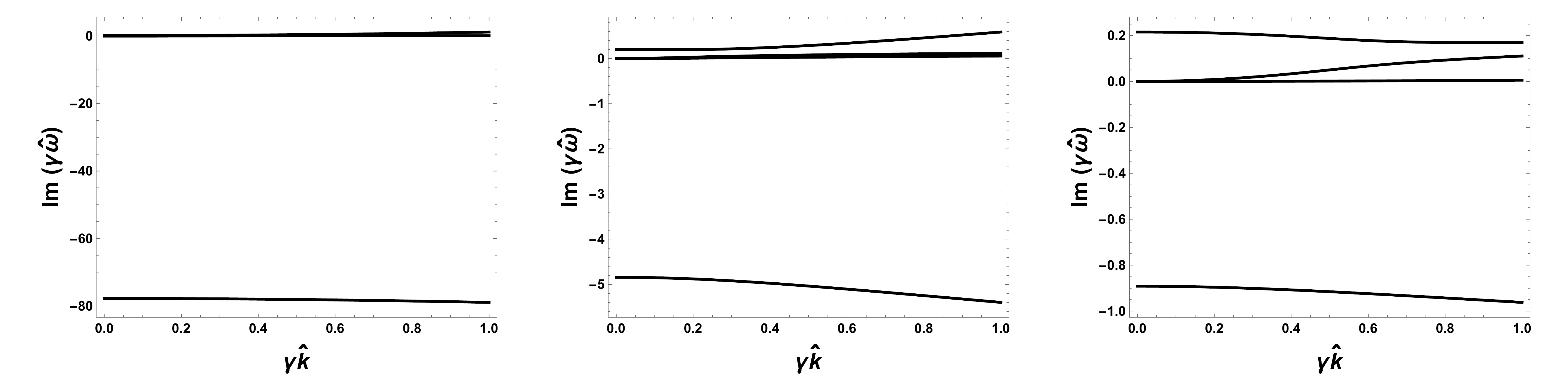}\\
\includegraphics[width=\textwidth]{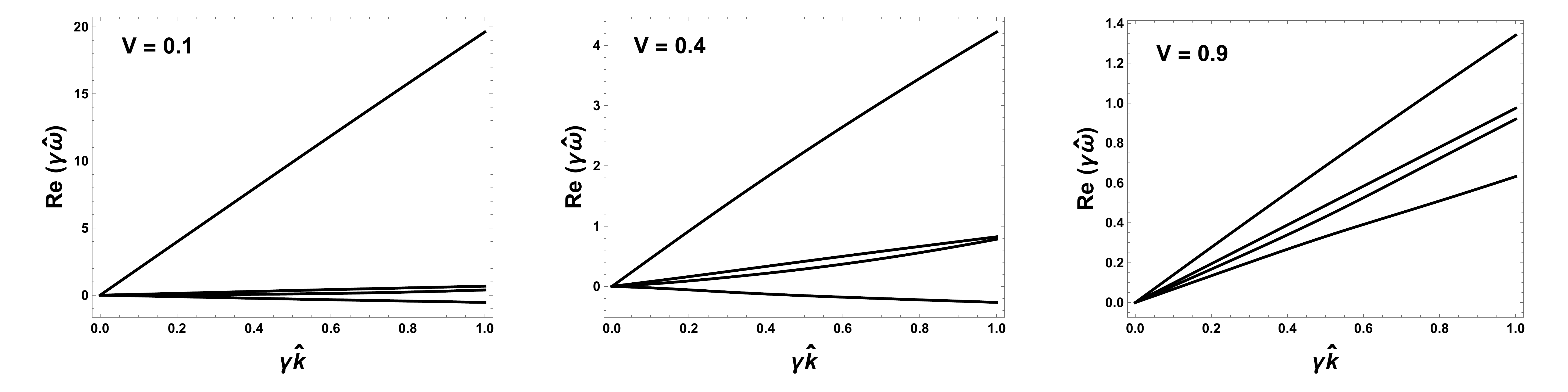}
\caption{Imaginary and real parts of the longitudinal modes, considering $\hat\tau_\pi=5$ \cite{dnmr}, for three different values of background velocity $V=0.1$, $V=0.4$, and $V=0.9$, in the ultrarelativistic limit $c_s^2=1/3$.}
\label{fig_nonstatic_long}
\end{center}
\end{figure}

The longitudinal modes for perturbations on a moving fluid, solutions of Eq.~(\ref{non-static_long}), are displayed in Fig.~\ref{fig_nonstatic_long}, considering $\hat\tau_\pi=5$ \cite{dnmr}, for several values of background velocity in the ultrarelativistic limit, $c_{\mathrm{s}}^2=1/3$.

As it was expected from the analysis performed for the transverse modes, the new nonhydrodynamic mode that appears for perturbations on a moving fluid is unstable not only in the vanishing wave number limit, but also for any value of $k$. On the other hand, the modes that are already present for perturbations on a static background fluid remain stable for any value of wave number and background velocity. 

The aforementioned instability is similar to what is observed in relativistic Navier-Stokes theory. In this context, it is intuitive to employ an approach analogous to the one proposed by Maxwell--Cattaneo \cite{maxwell, cattaneo} in order to obtain a linearly causal and stable equation of motion for gradients of shear-stress tensor. In this scenario, one would obtain an alternative third-order formulation in which these problems should no be longer present. In the next section we analyze how is it possible to modify this theory and the implications on its linear stability. 

\section{Modified third-order fluid dynamics}
\label{sec_hyper}

In the last section, the third-order fluid-dynamical formulation proposed in Ref.~\cite{amaresh} was shown to be linearly unstable. In this section, we propose a modified version of this theory that can be constructed to be linearly causal and stable. Note that the third-order equation of motion for shear-stress tensor was originally written in Eq.~\eqref{eq_amaresh_dots} as
\begin{equation}
\dot{\pi}^{\langle\mu\nu\rangle}=2\frac{\eta}{\tau_\pi}\sigma^{\mu\nu}-\frac{1}{\tau_\pi}\pi^{\mu\nu}+\frac{4}{35}\nabla^{\langle\mu}\left(\tau_\pi\nabla_\alpha \pi^{\nu\rangle\alpha}\right)-\frac{2}{7}\nabla_{\alpha}\left(\tau_\pi \nabla^{\langle\mu}\pi^{\nu\rangle\alpha}\right)-\frac{1}{7}\nabla_\alpha\left(\tau_\pi\nabla^{\alpha}\pi^{\langle\mu\nu\rangle}\right)+\cdots, \label{relax_shear}
\end{equation}
with the dots denoting contribution of nonlinear terms. The main reason for the occurrence of additional \textit{unstable} modes for perturbations on a moving fluid are the second-order spacelike derivatives of the shear-stress tensor on the right-hand side (while the equation only contains first-order timelike derivatives). If one then performs a Lorentz boost, such second-order spatial derivatives will lead to the appearance of second-order time derivatives and modify fundamental mathematical properties of the partial differential equation. In Fourier space, this manifests as the appearance of additional modes in the theory. 

One way to remove this problem is to convert all gradients of the shear-stress tensor into an independent dynamical variable
\begin{equation}
\nabla^{\langle\alpha}\pi^{\mu\nu\rangle}\longrightarrow\rho^{\alpha\mu\nu},\label{rho_amaresh}
\end{equation}
with the brackets denoting the contraction with a triple symmetric traceless projection operator onto the space orthogonal to the 4-velocity, $\nabla^{\langle\mu}\pi^{\nu\lambda\rangle}\equiv\Delta^{\mu\nu\lambda}_{\alpha\beta\gamma}\nabla^{\langle\alpha}\pi^{\beta\gamma\rangle}$. This projection operator is a sixth-rank tensor, defined as follows
\begin{eqnarray}
\Delta^{\mu\nu\lambda}_{\alpha\beta\rho}&\equiv&\frac{1}{6}\left[\Delta^\mu_\alpha\left(\Delta^\nu_\beta \Delta^\lambda_\rho+\Delta^\nu_\rho \Delta^\lambda_\beta\right)+\Delta^\mu_\beta\left(\Delta^\nu_\alpha \Delta^\lambda_\rho+\Delta^\nu_\rho \Delta^\lambda_\alpha\right)+\Delta^\mu_\rho\left(\Delta^\nu_\alpha \Delta^\lambda_\beta+\Delta^\nu_\beta \Delta^\lambda_\alpha\right)\right]\notag \\
&-&\frac{1}{15}\left[\Delta^{\mu\nu}\left(\Delta^\lambda_\alpha \Delta_{\beta\rho}+\Delta^\lambda_\beta \Delta_{\alpha\rho}+\Delta^\lambda_\rho \Delta_{\alpha\beta} \right)+\Delta^{\mu\lambda}\left(\Delta^\nu_\alpha \Delta_{\beta\rho}+\Delta^\nu_\beta \Delta_{\alpha\rho}+\Delta^\nu_\rho \Delta_{\alpha\beta}\right)\right .\notag \\
&+&\left .\Delta^{\nu\lambda}\left(\Delta^\mu_\alpha \Delta_{\beta\rho}+\Delta^\mu_\beta \Delta_{\alpha\rho}+\Delta^\mu_\rho \Delta_{\alpha\beta}\right)\right].
\end{eqnarray}
In this case, Eq.~(\ref{relax_shear}) is rewritten in the following form
\begin{equation}
\tau_\pi\dot{\pi}^{\langle\mu\nu\rangle}+\pi^{\langle\mu\nu\rangle}=2\eta\sigma^{\mu\nu}-\tau_\pi\nabla_\alpha\rho^{\alpha\mu\nu}+\cdots,\label{relax_shear2}
\end{equation}
where the dots once again denote all possible nonlinear terms that do not contribute to a linear stability analysis.

The next step is to impose that $\rho^{\mu\nu\lambda}$ is not simply proportional to gradients of the shear-stress tensor, but instead relaxes to such quantities exponentially,
\begin{equation}
\tau_\rho \dot{\rho}^{\langle\mu\nu\lambda\rangle}+\rho^{\mu\nu\lambda}=\frac{3}{7}\eta_\rho \nabla^{\langle\mu}\pi^{\nu\lambda\rangle}+\text{nonlinear terms},\label{rho_new}
\end{equation}
with $\tau_\rho$ being introduced as a novel relaxation time and $\eta_\rho$ as an effective viscosity coefficient associated with the new hydrodynamic variable $\rho^{\mu\nu\lambda}$. Note that if we take $\tau_\rho\rightarrow0$ and $\eta_\rho\longrightarrow\tau_\pi$, we recover the original version of the theory in the linear regime -- in particular, the factor $3/7$ in the right-hand side of Eq.~\eqref{rho_new} is essential to ensure this. Once again, we remark that a complete nonlinear theory will not be formally derived in this work, and we shall focus on the causality and stability of the novel formulation in the linear regime. Finally, we also remark that an analogous equation of motion for $\rho^{\mu\nu\lambda}$ can be derived from the Boltzmann equation, see Ref.~\cite{dnmr}. Above, we have assumed the simplest form $\eta_\rho$ can have -- one could also assume it to be a rank-six tensor.

Next, we consider a system initially in a global equilibrium state and proceed to perform small perturbations around such state. In this case, we must extend the previous linear stability analysis to also consider perturbations in $\rho^{\mu\nu\lambda}$,
\begin{equation}
\varepsilon=\varepsilon_0+\delta\varepsilon,\hspace{0.2cm}u^\mu=u^{\mu}_0+\delta u^{\mu},\hspace{0.2cm}\pi^{\mu\nu}=\delta\pi^{\mu\nu},\hspace{0.2cm}\rho^{\mu\nu\lambda}=\delta\rho^{\mu\nu\lambda}. \label{perturbs_hyper}
\end{equation}
In this case, the linearized Eqs.~(\ref{relax_shear2}) and (\ref{rho_new}) become
\begin{eqnarray}
\tau_\pi D_0 \delta\pi^{\mu\nu}+\delta\pi^{\mu\nu}&=&\eta\left(\nabla_0^\mu\delta u^\nu+\nabla_0^\nu\delta u^\mu-\frac{2}{3}\Delta_0^{\mu\nu}\partial_\lambda \delta u^\lambda\right)-\tau_\pi\nabla^0_\alpha\delta\rho^{\alpha\mu\nu}, \label{linear_pi}\\
\tau_\rho D_0\delta\rho^{\mu\nu\lambda}+\delta\rho^{\mu\nu\lambda}&=&\eta_\rho\left[\frac{1}{7}\left(\nabla_0^\lambda \delta\pi^{\mu\nu}+\nabla_0^\nu \delta\pi^{\mu\lambda}+\nabla_0^\mu \delta\pi^{\nu\lambda}\right)+\right .\notag \\
&-&\left . \frac{2}{35}\left(\Delta_0^{\mu\nu}\nabla^0_\alpha\delta\pi^{\lambda\alpha}+\Delta_0^{\mu\lambda}\nabla^0_\alpha\delta\pi^{\nu\alpha}+\Delta_0^{\nu\lambda}\nabla^0_\alpha\delta\pi^{\mu\alpha}\right)\right]. \label{linear_rho}
\end{eqnarray}
The next step is to calculate the Fourier transform of Eqs.~(\ref{linear_pi}) and (\ref{linear_rho}), which read
\begin{eqnarray}
\left(i\Omega\tau_\pi+1\right)\delta\tilde{\pi}^{\mu\nu}&=&i\eta\left(\kappa^\mu\delta\tilde{u}^\nu+\kappa^\nu\delta\tilde{u}^\mu-\frac{2}{3}\Delta^{\mu\nu}\kappa_\lambda \delta\tilde{u}^{\lambda}\right)-i\tau_\pi\kappa_\alpha\delta\tilde{\rho}^{\alpha\mu\nu}, \label{fourier_pi}\\
\left(i\Omega\tau_\rho+1\right)\delta\tilde{\rho}^{\mu\nu\lambda}&=&i\eta_\rho\left[\frac{1}{7}\left(\kappa^\lambda \delta\tilde{\pi}^{\mu\nu}+\kappa^\nu \delta\tilde{\pi}^{\mu\lambda}+\kappa^\mu \delta\tilde{\pi}^{\nu\lambda}\right)+\right .\notag \\
&-&\left . \frac{2}{35}\left(\Delta^{\mu\nu}\kappa_\alpha\delta\tilde{\pi}^{\lambda\alpha}+\Delta^{\mu\lambda}\kappa_\alpha\delta\tilde{\pi}^{\nu\alpha}+\Delta^{\nu\lambda}\kappa_\alpha\delta\tilde{\pi}^{\mu\alpha}\right)\right]. \label{fourier_rho}
\end{eqnarray}
Note that, on the right-hand side of Eq.~(\ref{fourier_pi}), only the projection $\kappa_\alpha\delta\tilde{\rho}^{\alpha\mu\nu}$ appears. In order to obtain this projection, we must contract Eq.~\eqref{fourier_rho} with $\kappa_{\mu}$,
\begin{eqnarray}
\left(i\Omega\tau_\rho+1\right)\kappa_\mu\delta\tilde{\rho}^{\mu\nu\lambda}&=&-\frac{i}{7}\eta_\rho\kappa^2\delta\tilde{\pi}^{\nu\lambda}+\frac{3i}{35}\eta_\rho\left(\kappa_\alpha \kappa^\nu \delta\tilde{\pi}^{\lambda\alpha}+\kappa_\alpha \kappa^\lambda \delta\tilde{\pi}^{\nu\alpha}\right)-\frac{2i}{35}\eta_\rho \Delta^{\nu\lambda}\kappa_\alpha \kappa_\beta \delta\tilde{\pi}^{\alpha\beta}. \label{rho_fourier}
\end{eqnarray}
As it was first performed in the last section, the linear stability analysis of the novel third-order theory shall be divided in the study of its transverse and longitudinal modes, employing the procedure first developed in Ref.~\cite{bd}.

\subsection{Transverse modes}

For the sake of consistency, we shall begin looking at the transverse modes of the novel formulation. First, we compute the transverse component of Eq.~(\ref{rho_fourier}), which is obtained by the following projection
\begin{equation}
\left(-\frac{\kappa_\mu}{\kappa}\Delta^\alpha_{\nu,\kappa}\right)\kappa_\lambda\delta\tilde{\rho}^{\mu\nu\lambda}=-\frac{8i}{35}\frac{\eta_\rho \kappa^2}{i\Omega\tau_\rho+1}\delta\tilde{\pi}^\alpha_\bot.
\end{equation}
Therefore, inserting this equation in the partially transverse projection of Eq.~(\ref{fourier_pi}), we obtain
\begin{equation}
\left(i\hat{\tau}_{\pi}\hat{\Omega}+\frac{8}{35}\frac{\hat\eta_\rho\hat{\tau}_{\pi}\hat{\kappa}^2}{i\hat\Omega \hat\tau_\rho+1}+1\right)\frac{\delta\tilde{\pi}^\mu_{\bot}}{\varepsilon_0+P_0}-i\hat{\kappa}\delta\tilde{u}^\mu_\bot=0.
\end{equation}
Naturally, the transverse projection of Eq.~(\ref{fourier2}) remains unchanged and is given by Eq.~(\ref{eom_trans1}). Therefore, the equations that describe the transverse degrees of freedom of the modified third-order theory can be written in a matrix form as
\begin{equation}
\left( 
\begin{array}{cc}
i\hat{\tau}_{\pi}\hat{\Omega}+\frac{8}{35}\frac{\hat\eta_\rho\hat{\tau}_{\pi}\hat{\kappa}^2}{i\hat\Omega \hat\tau_\rho+1}+1 & -i\hat{\kappa}  \\ 
-\hat{\kappa} & \hat{\Omega} 
\end{array}%
\right) \left( 
\begin{array}{c}
\frac{\delta\tilde{\pi}^\mu_{\bot}}{\varepsilon_0+P_0} \\
\delta \hat{u}_{\bot}^{\mu }
\end{array}%
\right)=0,
\end{equation}
In this case, the dispersion relation associated with the transverse modes read
\begin{equation}
\hat{\Omega}\left(i\hat{\tau}_{\pi}\hat{\Omega}+\frac{8}{35}\frac{\hat{\tau}_{\pi} \hat\eta_\rho\hat{\kappa}^2}{i\hat\Omega \hat{\tau}_\rho+1}+1\right)-i\hat{\kappa}^2=0.\label{disp_trans_new}
\end{equation}
One can straightforwardly recover the dispersion relation associated with the transverse modes of the original third-order formulation by simply taking $\hat\eta_\rho=\hat\tau_\pi$ and $\hat\tau_\rho=0$, see Eq.~(\ref{disp_trans}). However, in the last section, we showed that such theory yields acausal and unstable modes for perturbations on a moving fluid. In this section, we will analyze whether the inclusion of the transient dynamics of a hydrodynamic current defined as $\rho^{\mu\nu\lambda}$ is sufficient to render the modified version of the third-order theory linearly causal and stable.

\subsubsection{Perturbations on a static fluid}

Once again, we begin looking at the transverse modes for perturbations on a static background fluid, $V=0$. In this case, the dispersion relation, Eq.~(\ref{disp_trans_new}), reads simply

\begin{equation}
\hat{\omega}\left(i\hat{\tau}_{\pi}\hat{\omega}+\frac{8}{35}\frac{\hat{\tau}_{\pi} \hat\eta_\rho\hat k^2}{i\hat\omega \hat{\tau}_\rho+1}+1\right)-i\hat k^2=0. \label{trans_hyper_stat}
\end{equation}
As it was done in the previous analysis, we shall look at the asymptotic form of these modes. In the small wave number limit, these modes can be written as
\begin{eqnarray}
\hat\omega^{\mathrm{shear}}_{T,-}&=&i\hat k^2+i\hat{\tau}_{\pi}\left(1-\frac{8\hat\eta_\rho}{35}\right)\hat k^4+\mathcal{O}\left(\hat k^6\right),\\
\hat\omega^{\mathrm{shear}}_{T,+}&=&\frac{i}{\hat\tau_\pi}+i\left[1-\frac{8\hat\tau_\pi\hat\eta_\rho}{35\left(\hat\tau_\pi-\hat\tau_\rho\right)}\right]\hat k^2+\mathcal{O}\left(\hat k^4\right),\\
\hat\omega^{\mathrm{shear}}_{T,\mathrm{new}}&=&\frac{i}{\hat\tau_\rho}-\frac{8i\hat\tau_\pi \hat\eta_\rho}{35\left(\hat\tau_\pi-\hat\tau_\rho\right)}\hat k^2+\mathcal{O}\left(\hat k^4\right).
\end{eqnarray}
As expected, the inclusion of the nonconserved current defined as $\rho^{\mu\nu\lambda}$ leads to the occurrence of a new nonhydrodynamic mode already for perturbations on a static background fluid, in which there is an additional mode in comparison to the transverse modes of the previous theory. In particular, this new mode behaves as $\hat\omega\sim i/\hat\tau_\rho$ in the small wave number limit. Furthermore, note that we immediately recover the modes of the original third-order theory by taking $\tau_\rho\rightarrow0$ and $\eta_\rho\rightarrow\tau_\pi$.

In the large wave number limit, the transverse modes of this theory are
\begin{eqnarray}
\hat\omega^{\mathrm{shear}}_{T,-}&=&\frac{35i}{35\hat\tau_\rho+8\hat\tau_\pi \hat\eta_\rho}-9800i\hat\tau_\pi\hat\eta_\rho\frac{35\hat\tau_\rho+\hat\tau_\pi\left(8\hat\eta_\rho-35\right)}{\left(35\hat\tau_\rho+8\hat\tau_\pi \hat\eta_\rho\right)^4}\frac{1}{\hat k^2}+\mathcal{O}\left(\frac{1}{\hat k^4}\right),\\
\hat\omega^{\mathrm{shear}}_{T,+}&=&\hat k\sqrt{\frac{35\hat\tau_\rho+8\hat\tau_\pi\hat\eta_\rho}{35\hat\tau_\pi\hat\tau_\rho}}+i\frac{35\hat\tau_\rho^2+8\hat\tau_\pi^2 \hat\eta_\rho+8\hat\tau_\pi\hat\tau_\rho\hat\eta_\rho}{2\hat\tau_\pi\hat\tau_\rho\left(35\hat\tau_\rho+8\hat\tau_\pi\hat\tau_\rho\right)}+\mathcal{O}\left(\frac{1}{\hat k^2}\right),\\
\hat\omega^{\mathrm{shear}}_{T,\mathrm{new}}&=&-\hat k\sqrt{\frac{35\hat\tau_\rho+8\hat\tau_\pi\hat\eta_\rho}{35\hat\tau_\pi\hat\tau_\rho}}+i\frac{35\hat\tau_\rho^2+8\hat\tau_\pi^2 \hat\eta_\rho+8\hat\tau_\pi\hat\tau_\rho\hat\eta_\rho}{2\hat\tau_\pi\hat\tau_\rho\left(35\hat\tau_\rho+8\hat\tau_\pi\hat\tau_\rho\right)}+\mathcal{O}\left(\frac{1}{\hat k^2}\right).
\end{eqnarray}
Unlike what is observed for the third-order theory discussed in Sec.~\ref{sec_third}, in which the modes have a diffusionlike behavior in the large wave number limit, i.e., $\hat\omega\sim i\hat k^2$, this is not the case for the novel formulation. Furthermore, in the small wave number limit, the transverse modes are purely imaginary, while at large values of $\hat k$ they become propagating, i.e., their real parts are no longer zero. Therefore, it is now necessary to impose constraints on the linear causality of this theory. The asymptotic group velocity must be subluminal \cite{jackson}, and we thus obtain
\begin{equation}
\lim_{\hat k\rightarrow \infty}\left\vert\frac{\partial \mathrm{Re}(\hat\omega)}{\partial \hat k}\right\vert\leq 1\Longrightarrow \hat{\tau}_\rho\left(\hat\tau_\pi-1\right)\geq \frac{8}{35}\hat{\tau}_\pi\hat\eta_\rho. \label{causal_trans}
\end{equation}

\begin{figure}[ht]
\begin{center}
\includegraphics[width=.9\textwidth]{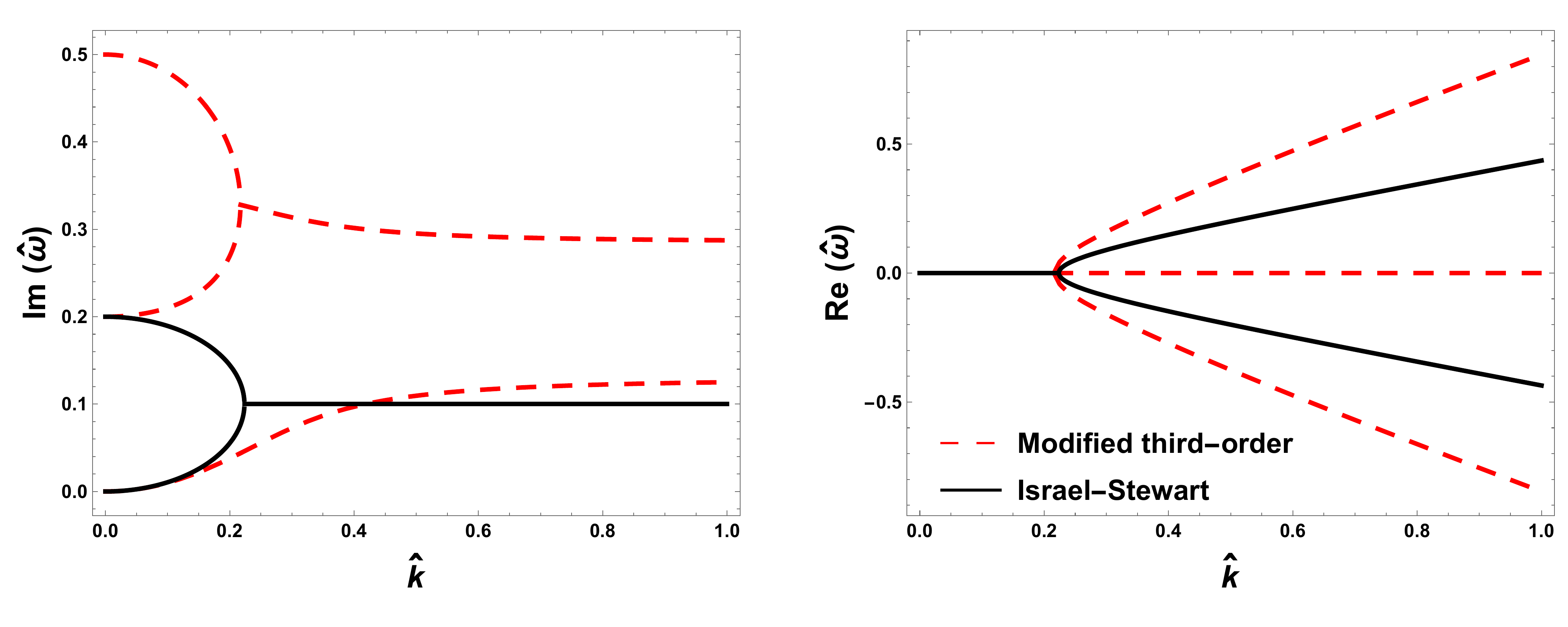}
\caption{Imaginary and real parts of the transverse modes of Israel-Stewart theory (solid black lines) and modified third-order fluid dynamics (red dashed lines) for perturbations on a static background fluid, considering $\hat{\tau}_\pi=\hat\eta_\rho=5$ \cite{dnmr} and $\hat\tau_\rho=2$.}
\label{is_hyper_trans}
\end{center}
\end{figure}

The solutions of Eq.~\eqref{trans_hyper_stat} are displayed in Fig.~\ref{is_hyper_trans} in comparison to the transverse modes of Israel-Stewart theory for perturbations on a static background fluid, assuming $\hat\tau_\pi=5$ \cite{dnmr} and $\hat\tau_\rho=2$. The novel third-order formulation has three transverse modes, while both the third-order formulation and Israel-Stewart theory have two transverse modes each. Considering the values for the transport coefficients employed here, the modes are stable not only in the vanishing wave number regime, but also for any value of $k$. Furthermore, while the transverse modes of the original third-order theory have vanishing real parts for perturbations on a static background fluid, the modes of the modified formulation have nonzero real parts, and thus are propagating.

\subsubsection{Perturbations on a moving fluid}

We now consider the case of perturbations on a moving fluid. Once again, we assume that the background fluid velocity is parallel to the wave vector. In this case, one must insert Eqs.~\eqref{omega} and \eqref{kappa} into the dispersion relation for the transverse modes, Eq.~(\ref{disp_trans_new}), leading to
\begin{equation}
\gamma(\hat{\omega}-V \hat k)\left[i\hat{\tau}_{\pi}\gamma(\hat{\omega}-V \hat k)+\frac{8}{35}\frac{\hat{\tau}_{\pi} \hat\eta_\rho\gamma^2(\hat{\omega}V-\hat k)^2}{i\gamma(\hat{\omega}-V \hat k)\hat{\tau}_\rho+1}+1\right]-i\gamma^2(\hat{\omega}V-\hat k)^2=0. \label{eq_hyper_trans_boost}
\end{equation}
Naturally, the solutions of this equation are rather complicated and shall not be written here. Instead, once again we resort to the analysis of the asymptotic form of these modes. In the small wave number limit, the modes can be written as
\begin{eqnarray}
\hat\omega^{\mathrm{shear}}_{T,-}&=&V\hat k+\frac{i}{\gamma}\hat k^2+\mathcal{O}\left(\hat k^3\right),\\
\hat\omega^{\mathrm{shear}}_{T,+}&=&\frac{i}{2\gamma}\left[\frac{\hat\tau_\pi+\hat\tau_\rho-V^2+\sqrt{(\hat\tau_\pi+\hat\tau_\rho-V^2)^2+\frac{32}{35}\hat\tau_\pi \hat\eta_\rho V^2+4\hat \tau_\rho\left(V^2-\hat\tau_\pi\right)}}{\hat\tau_\rho(\hat\tau_\pi-V^2)-\frac{8}{35}\hat\eta_\rho \hat\tau_\pi V^2}\right]+\mathcal{O}\left(\hat k\right),\\
\hat\omega^{\mathrm{shear}}_{T,\mathrm{new}}&=&\frac{i}{2\gamma}\left[\frac{\hat\tau_\pi+\hat\tau_\rho-V^2-\sqrt{(\hat\tau_\pi+\hat\tau_\rho-V^2)^2+\frac{32}{35}\hat\tau_\pi \hat\eta_\rho V^2+4\hat \tau_\rho\left(V^2-\hat\tau_\pi\right)}}{\hat\tau_\rho(\hat\tau_\pi-V^2)-\frac{8}{35}\hat\eta_\rho \hat\tau_\pi V^2}\right]+\mathcal{O}\left(\hat k\right).
\end{eqnarray}
As it was also observed for perturbations on a static fluid, there is one hydrodynamic mode, and two nonhydrodynamic modes. In particular, the hydrodynamic mode is identical to the one obtained for the third-order theory, see Eq.~\eqref{trans_mov1}. Furthermore, we recover the modes given by Eqs.~\eqref{trans_mov2} and (\ref{trans_mov3}) by simply taking $\hat\tau_\rho\rightarrow0$ and $\hat\eta_\rho\rightarrow\hat\tau_\pi$. We shall not study the behavior of these modes at large values of $k$, since the expansion for small wave number is enough to provide us constraints on the linear stability of this theory.

These modes are stable if their imaginary part is positive, which is guaranteed if both numerator and denominator have the same sign. Once again, we impose neither change their signs for any causal value of the background velocity $V$, otherwise resulting in a discontinuity in the modes. We note that both the denominator and the numerator are positive for $V=0$. Thus, they must remain positive for all causal values of velocity.

In order for the denominator to be positive for $0\leq V\leq1$, the transport coefficients must satisfy,
\begin{equation}
\hat\tau_\rho(\hat\tau_\pi-1)>\frac{8}{35}\hat\tau_\pi \hat\eta_\rho.\label{trho_trans}
\end{equation}
Note that this relation is identical to the linear causality condition obtained in Eq.~\eqref{causal_trans}. This relation also guarantees that if the term inside the square root is positive, it is smaller than the term outside it. Furthermore, in order for the numerator to be positive for all physical values of velocity, the relaxation times must satisfy the condition
\begin{equation}
\hat\tau_\pi+\hat\tau_\rho>1, \label{trans_cond2}
\end{equation}
which guarantees that the term outside the square root in the numerator is always positive. We note that this constraint reduces to the linear causality and stability conditions obtained for Israel-Stewart theory if $\tau_\rho=0$, see Refs.~\cite{bd, rischke}. However, we further remark that the stability condition given by Eq.~(\ref{trho_trans}) forbids this limit -- a linearly causal and stable theory can only be obtained if $\tau_\rho$ is not zero. Finally, note that Eqs.~(\ref{trho_trans}) and (\ref{trans_cond2}) combined guarantee that the square root in the numerator, if real, is always smaller than $\hat\tau_\pi+\hat\tau_\rho - V^2$, leading to a stable transverse mode. If the square root in the numerator is not real, then it does not contribute to the stability of the mode.

\begin{figure}[ht]
\begin{center}
\includegraphics[width=\textwidth]{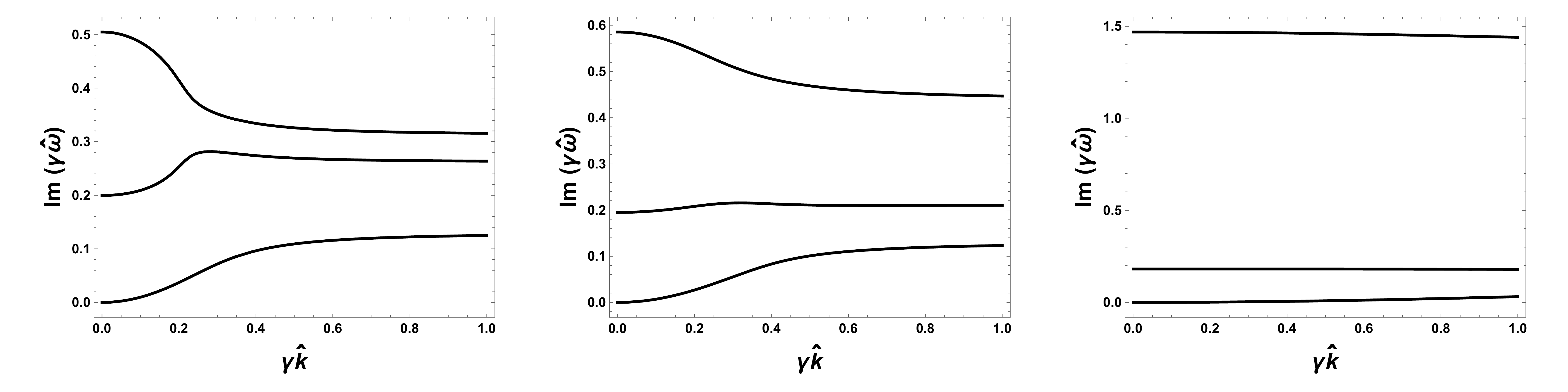}\\
\includegraphics[width=\textwidth]{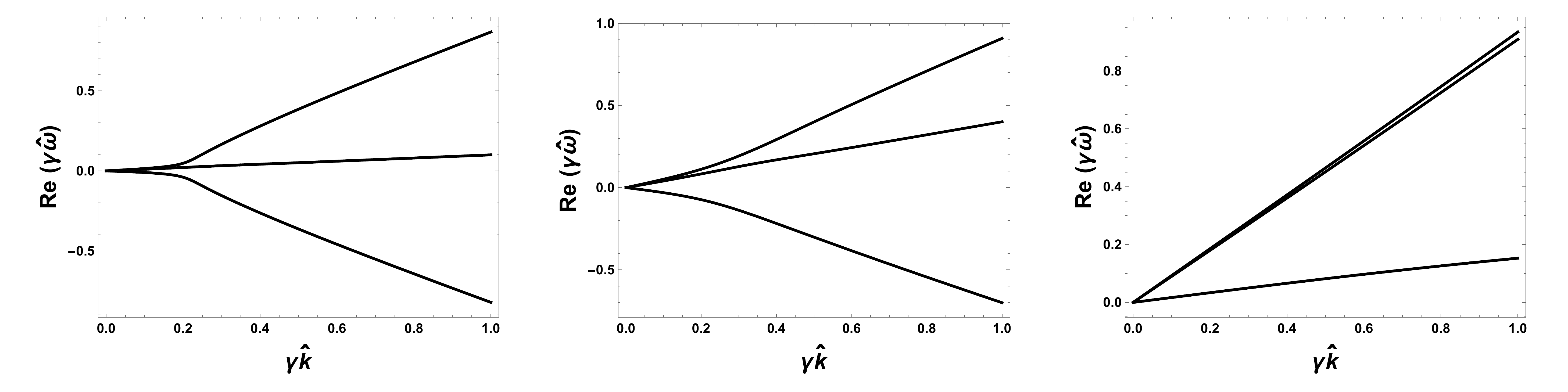}
\caption{Imaginary and real parts of the transverse modes for $\hat{\tau}_\pi=\hat\eta_\rho=5$, $\hat{\tau}_\rho=2$, considering three different values for the background velocity $V=0.1$, $V=0.4$, and $V=0.9$, respectively}
\label{fig_relax_nonstatic}
\end{center}
\end{figure}

For the sake of illustration, the solutions of Eq.~\eqref{eq_hyper_trans_boost} are displayed as function of wave number in Fig.~\ref{fig_relax_nonstatic}, considering $\hat\eta_\rho=\hat\tau_\pi=5$ \cite{dnmr} and $\hat\tau_\rho=2$, for three values for the background velocity and transport coefficients that satisfy the linear stability conditions for the transverse modes, Eqs.~(\ref{trho_trans}) and (\ref{trans_cond2}). Here, one can see that, for perturbations on a moving background fluid, the imaginary part of the nonhydrodynamic modes (upper panels) no longer merge for large values of the wave number.

The analysis developed so far in this section is restricted to the transverse modes of the modified third-order theory. Even though we obtained the conditions the novel transport coefficients must satisfy in order to guarantee the theory's linear causality and stability, it is still necessary to look at the conditions that arise from the longitudinal modes.

\subsection{Longitudinal modes}

Naturally, the longitudinal projections of Eqs.~(\ref{fourier1}) and (\ref{fourier2}) remain the same, as given by Eqs.~(\ref{eom_long1}) and (\ref{eom_long2}), respectively. Furthermore, the longitudinal component of Eq.~(\ref{rho_fourier}) is obtained by contracting it with $\kappa_\mu \kappa_\nu$, thus leading to
\begin{equation}
\left(i\Omega\tau_\rho+1\right)\left(\frac{\kappa_\mu \kappa_\nu}{\kappa^2}\right)\kappa_\lambda \delta\tilde{\rho}^{\mu\nu\lambda}=-\frac{9i}{35}\eta_\rho \kappa^2 \delta\tilde{\pi}_\|.
\end{equation}
Inserting this result in the longitudinal projection of Eq.~(\ref{fourier_pi}), we obtain
\begin{equation}
\left(i\hat{\tau}_{\pi}\hat{\Omega}+\frac{9}{35}\frac{\hat\eta_\rho \hat{\tau}_{\pi}\hat{\kappa}^2}{i\hat\Omega \hat\tau_\rho+1}+1\right)\frac{\delta\tilde{\pi}_{\|}}{\varepsilon_0+P_0}-\frac{4i}{3}\hat{\kappa}\delta\tilde{u}_\|=0.
\end{equation}
It is possible to write the equation for the longitudinal modes in the following matrix form
\begin{equation}
\left( 
\begin{array}{ccc}
i\hat{\tau}_{\pi}\hat{\Omega}+\frac{9}{35}\frac{\hat\eta_\rho \hat{\tau}_{\pi}\hat{\kappa}^2}{i\hat\Omega \hat\tau_\rho+1}+1 & -i\frac{4}{3}\hat{\kappa} & 0  \\ 
-\hat{\kappa} & \hat{\Omega} & -c_{\mathrm{s}}^2 \hat{\kappa} \\
0 & -\hat{\kappa} & \hat{\Omega}
\end{array}%
\right) \left( 
\begin{array}{c}
\frac{\delta\tilde{\pi}_{\|}}{\varepsilon_0+P_0} \\ 
\delta \tilde{u}_{\|} \\
\frac{\delta\varepsilon}{\varepsilon_0+P_0}
\end{array}%
\right)=0.
\end{equation}
Therefore, the dispersion related to the longitudinal degrees of freedom of the novel third-order formulation reads
\begin{equation}
\left(\hat{\Omega}^2-c_s^2\hat{\kappa}^2\right)\left(i\hat{\tau}_{\pi}\hat{\Omega}+\frac{9}{35}\frac{\hat{\tau}_{\pi} \hat\eta_\rho \hat{\kappa}^2}{i\hat\Omega \hat\tau_\rho+1}+1\right)-\frac{4i}{3}\hat{\Omega}\hat{\kappa}^2=0.\label{disp_long_new}
\end{equation}
Once again, one can straightforwardly recover the dispersion relation for the longitudinal modes using the formulation developed in Ref.~\cite{amaresh}, see Eq.~(\ref{disp_long}), by simply taking the novel relaxation time to zero, $\tau_\rho=0$, and $\eta_\rho=\tau_\pi$. As it was also observed for the transverse modes of this theory, the dispersion relation associated with the longitudinal modes is an one degree higher polynomial when compared to the original third-order theory. This is expected, since including the dynamics of a nonconserved hydrodynamic current $\rho^{\mu\nu\lambda}$ leads to the occurrence of an additional nonhydrodynamic mode. However, unlike what we observed for the original third-order theory, the number of modes of the modified theory does not increase when considering perturbations on a moving fluid.

\subsubsection{Perturbations on a static fluid}

We first look at the longitudinal modes of the theory for perturbations on a static fluid. In this case, the dispersion relation associated with the longitudinal modes, Eq.~(\ref{disp_long_new}), simply reads
\begin{equation}
\left(\hat{\omega}^2-c_s^2\hat{k}^2\right)\left(i\hat{\tau}_{\pi}\hat{\omega}+\frac{9}{35}\frac{\hat{\tau}_{\pi} \hat\eta_\rho \hat{k}^2}{i\hat\omega \hat\tau_\rho+1}+1\right)-\frac{4i}{3}\hat{\omega}\hat{k}^2=0.
\end{equation}
As before, let us analyze the asymptotic form of the modes. In the small wave number limit, they read
\begin{eqnarray}
\hat\omega^{\mathrm{sound}}_\pm&=&\pm c_\mathrm{s}\hat k+\frac{2i}{3}\hat k^2\pm\frac{2\left(3\hat\tau_\pi c_\mathrm{s}^2-1\right)}{9 c_\mathrm{s}}\hat k^3+\mathcal{O}\left(\hat k^4\right), \label{hyper_sound}\\
\hat\omega^{\mathrm{shear}}_{L}&=&\frac{i}{\hat\tau_\pi}+ i\frac{\frac{27}{35}\hat\tau_\pi\hat\eta_\rho+4(\hat\tau_\rho-\hat\tau_\pi)}{3(\hat\tau_\pi-\hat\tau_\rho)}\hat k^2+\mathcal{O}\left(\hat k^4\right), \label{hyper_shear_L}\\
\hat\omega^{\mathrm{shear}}_{L,\mathrm{new}}&=&\frac{i}{\hat\tau_\rho}- i\frac{9\hat\tau_\pi\hat\eta_\rho}{35(\hat\tau_\pi-\hat\tau_\rho)}\hat k^2+\mathcal{O}\left(\hat k^4\right). \label{hyper_shear_new}
\end{eqnarray}
We identify two hydrodynamic modes and two nonhydrodynamic modes. The hydrodynamic modes correspond to the usual sound modes and their small wave number limit remain identical to the corresponding results obtained in Israel-Stewart theory or in the original version of the third-order theory, see Eq.~\eqref{parabolic_long_stat1}. In the small wave number limit, the nonhydrodynamic mode
$\hat\omega^{\mathrm{shear}}_{L}$ becomes identical to the nonhydrodynamic mode found in Israel-Stewart theory and in the original third-order theory, with deviations only occurring at order $\mathcal{O}(\hat{k}^2)$, see Eq.~\eqref{shear_L_para}.
On the other hand, the nonhydrodynamic mode $\hat\omega^{\mathrm{shear}}_{L, \mathrm{new}}$ is intrinsically new and describes nonequilibrium modes that relax to equilibrium in times of order $\tau_\rho$. We note that, as expected, we recover the results from the previous section, Eq.~\eqref{shear_L_para}, when taking $\hat\tau_\rho\rightarrow0$ and $\hat\eta_\rho\rightarrow\hat\tau_\pi$.

In the large wave number limit, all four longitudinal modes can be cast in the following form
\begin{eqnarray}
\hat\omega&=&\pm\sqrt{\frac{\frac{27}{35}\hat\tau_\pi \hat\eta_\rho+3\hat\tau_\pi\hat\tau_\rho c_{\mathrm{s}}^2+4\hat\tau_\rho\pm\sqrt{\left(\frac{27}{35}\hat\tau_\pi \hat\eta_\rho+3\hat\tau_\pi\hat\tau_\rho c_{\mathrm{s}}^2+4\hat\tau_\rho\right)^2-\frac{324}{35}\hat\tau_\pi^2 \hat\tau_\rho \hat\eta_\rho c_{\mathrm{s}}^2}}{6\hat\tau_\pi\hat\tau_\rho}}\hat k+\mathcal{O}\left(1\right).
\end{eqnarray}
Since the hydrodynamic and nonhydrodynamic modes merge at finite values of wave number, it is not trivial to map these four solutions with the small wave number solutions displayed in Eqs.~\eqref{hyper_sound}, \eqref{hyper_shear_L}, and \eqref{hyper_shear_new}. In order for these modes to be stable, it is essential that the term inside the outer square root is real and positive, otherwise leading to modes with a negative imaginary part, and thus unstable solutions. For this purpose, we must first impose that the term inside the inner square root is positive. If this is the case, it is straightforward to see that the term in the numerator is always positive. Therefore, in order to obtain purely real modes, it is sufficient to impose
\begin{equation}
\left(\frac{27}{35}\hat\tau_\pi \hat\eta_\rho+3\hat\tau_\pi\hat\tau_\rho c_{\mathrm{s}}^2+4\hat\tau_\rho\right)^2-\frac{324}{35}\hat\tau_\pi^2 \hat\tau_\rho \hat\eta_\rho c_{\mathrm{s}}^2\geq0.
\end{equation}
In fact, this inequality is satisfied as long as the transport coefficients are positive definite quantities, i.e., $\hat\tau_\pi>0$, $\hat\tau_\rho>0$, and $\hat\eta_\rho>0$. Therefore, the stability of the longitudinal modes perturbations on a static fluid is always fulfilled. 

The linear causality of the theory can be verified by analyzing the asymptotic group velocity of the modes \cite{jackson}. In order for these modes to propagate subluminally, the following condition must be satisfied
\begin{equation}
\lim_{\hat k\rightarrow \infty}\left\vert\frac{\partial \mathrm{Re}(\hat\omega)}{\partial \hat k}\right\vert\leq 1\Longrightarrow\hat\tau_\rho\geq\frac{27}{35}\hat\eta_\rho \hat\tau_\pi\frac{1-c_{\mathrm{s}}^2}{3\hat\tau_\pi\left(1-c_{\mathrm{s}}^2\right)-4}. \label{causal_hyper_long}
\end{equation}
In order to obtain this relation, it is necessary to impose a first constraint to the shear relaxation time $\hat\tau_\pi$, which is given by
\begin{equation}
\hat\tau_\pi\geq\frac{4}{3(1-c_{\mathrm{s}}^2)},
\end{equation}
which, in the ultrarelativistic limit, $c_{\mathrm{s}}^2=1/3$, reduces to $\hat\tau_\pi\geq2$. This is exactly the linear stability condition for the shear relaxation time in Israel-Stewart theory \cite{rischke, bd}.

\begin{figure}[ht]
\begin{center}
\includegraphics[width=.9\textwidth]{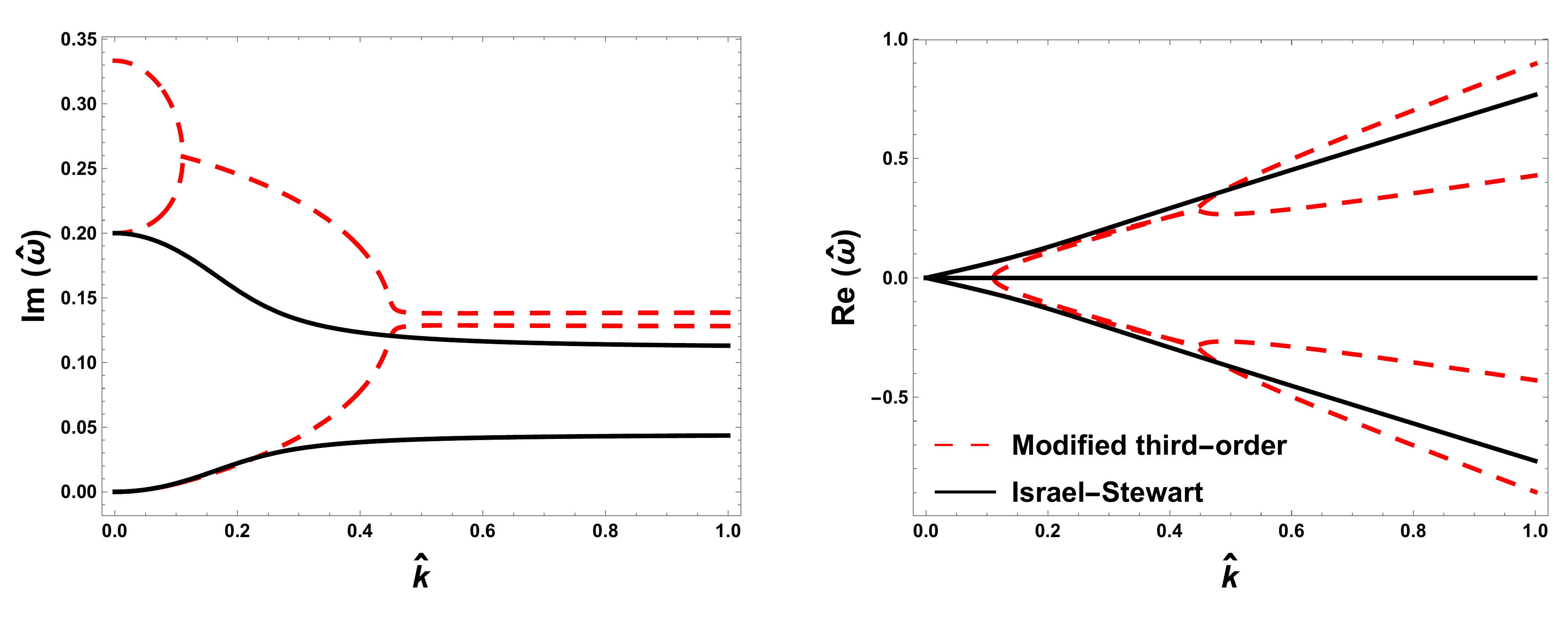}
\caption{Imaginary and real parts of the longitudinal modes of Israel-Stewart theory (solid black lines) and modified third-order fluid dynamics (red dashed lines) for perturbations on a static background fluid, considering $\hat{\tau}_\pi=\hat\eta_\rho=5$ \cite{dnmr} and $\hat\tau_\rho=3$ in the ultrarelativistic regime, $c_\mathrm{s}^2=1/3$.}
\label{is_hyper_long}
\end{center}
\end{figure}

The longitudinal modes of the modified third-order theory for perturbations on a static fluid are displayed in Fig.~\ref{is_hyper_long} in comparison with the corresponding longitudinal modes of Israel-Stewart theory for a wide range of values of wave number, considering $\hat\eta_\rho=\hat\tau_\pi=5$ \cite{dnmr} and $\hat\tau_\rho=3$.

\subsubsection{Perturbations on a moving fluid}

Next, we consider the longitudinal modes for perturbations on a moving fluid. Once again, we assume that the background velocity is parallel to the wave vector, thus $\hat\Omega$ and $\hat\kappa$ are given by Eqs.~(\ref{omega}) and (\ref{kappa}). In this case, the dispersion relation associated with the longitudinal modes, Eq.~(\ref{disp_long_new}), can be written as
\begin{equation}
\gamma^2\left[(\hat{\omega}-V\hat k)^2-c_s^2(\hat\omega V-\hat k)^2\right]\left[i\hat{\tau}_{\pi}\gamma(\hat\omega-V \hat k)+\frac{9}{35}\frac{\hat{\tau}_{\pi} \hat\eta_\rho \gamma^2(\hat\omega V-\hat k)^2}{i\hat\tau_\rho\gamma(\hat\omega-V \hat k)+1}+1\right]-\frac{4i}{3}\gamma^3(\hat\omega-V\hat k)(\hat\omega V-\hat k)^2=0. \label{eq_hyper_long_boost}
\end{equation}
Once again, the general solutions for this case can be extremely complicated and will not be displayed here. Instead, we analyze the asymptotic form of these modes. First, in the small wave number limit, they read
\begin{eqnarray}
\hat\omega^{\mathrm{sound}}_\pm&=&\frac{V\pm c_{\mathrm{s}}}{1\pm c_\mathrm{s}V}\hat k+\mathcal{O}\left(\hat k^2\right),\\
\hat\omega^{\mathrm{shear}}_L&=&-i\frac{3\mathcal{A}\mathcal{C}-4V^2-\sqrt{(3\mathcal{A}\mathcal{C}-4V^2)^2+12\mathcal{A}\left(\mathcal{A}\mathcal{B}+4\hat\tau_\rho V^2\right)}}{2\gamma\left(\mathcal{A}\mathcal{B}+4\hat\tau_\rho V^2\right)}+\mathcal{O}\left(\hat k\right),\\
\hat\omega^{\mathrm{shear}}_{L,\mathrm{new}}&=&-i\frac{3\mathcal{A}\mathcal{C}-4V^2+\sqrt{(3\mathcal{A}\mathcal{C}-4V^2)^2+12\mathcal{A}\left(\mathcal{A}\mathcal{B}+4\hat\tau_\rho V^2\right)}}{2\gamma\left(\mathcal{A}\mathcal{B}+4\hat\tau_\rho V^2\right)}+\mathcal{O}\left(\hat k\right),
\end{eqnarray}
where the following variables were introduced
\begin{eqnarray}
\mathcal{A}&\equiv&1-\mathrm{c}_\mathrm{s}^2V^2,\\
\mathcal{B}&\equiv&3\hat\tau_\pi\left(\frac{9}{35}\hat\eta_\rho V^2-\hat\tau_\rho\right),\\
\mathcal{C}&\equiv&\hat\tau_\pi+\hat\tau_\rho.
\end{eqnarray}
Since we are interested in obtaining linear stability conditions imposed by the longitudinal modes, it is sufficient to solely analyze the modes for $V\neq0$ in the small wave number limit, and thus we shall not study these modes for large wave number in this work.

There are two hydrodynamic longitudinal modes, which are purely real and thus always stable in this regime, and two nonhydrodynamic longitudinal modes. As it was observed for the original third-order theory, analyzed in the previous section, the hydrodynamic modes for perturbations on top of a moving background are given by the relativistic velocity addition, see Eq.~\eqref{long_rel_vel}. The stability of the nonhydrodynamic modes is not automatically guaranteed and must be analyzed in more detail. These modes are stable if both the numerator and denominator have opposite signs, leading the modes to have a positive imaginary part. Once again, we make the assumption that neither the numerator nor the denominator change their signs for any causal value of the background velocity $V$, otherwise leading to a problematic discontinuity. Taking the background fluid velocity to be zero, $V=0$, one can see that the numerator is positive definite, while the denominator is negative. In general, the denominator is negative for any value the background fluid velocity can assume as long as the following condition is satisfied
\begin{equation}
\mathcal{A}\mathcal{B}+4\hat\tau_\rho V^2<0,
\end{equation}
which leads to the following conditions
\begin{eqnarray}
\hat\tau_\rho&>&\frac{27}{35}\hat\eta_\rho \hat\tau_\pi\frac{1-c_{\mathrm{s}}^2}{3\hat\tau_\pi\left(1-c_{\mathrm{s}}^2\right)-4}, \label{trho_long}\\
\hat\tau_\pi&>&\frac{4}{3(1-c_{\mathrm{s}}^2)}. \label{taupi_long_final}
\end{eqnarray}
Note that the constraint given by Eq.~\eqref{trho_long} is identical to the linear causality condition obtained when analyzing the modes for perturbations on a static fluid in the large wave number limit, see Eq.~\eqref{causal_hyper_long}. Furthermore, in the ultrarelativistic limit, $c_{\mathrm{s}}^2=1/3$, Eq.~\eqref{taupi_long_final} reduces to $\hat\tau_\pi>2$, which corresponds to the linear causality and stability condition for the shear relaxation time in Israel-Stewart theory \cite{rischke, bd}.

The final step is to evaluate the necessary condition to obtain a positive numerator. First, we note that, in order to obtain linearly stable modes, the term inside the square root in the numerator must be either: positive \textit{and} smaller than the term outside, or negative. Both conditions are guaranteed by imposing $3\mathcal{A}\mathcal{C}-4V^2\geq0$. Naturally, this constraint must be valid for any physical value of the fluid velocity, $V$. In this case, the strongest condition possible is obtained considering the maximum value for the background velocity, $V=1$. Then, we have
\begin{equation}
3(1-\mathrm{c}_\mathrm{s}^2)(\hat\tau_\pi+\hat\tau_\rho)\geq4, \label{rho+pi_long}
\end{equation}
which is guaranteed by Eq.~\eqref{taupi_long_final}.

Note that the stability condition given by Eq.~\eqref{rho+pi_long} reduces to the linear causality and stability conditions derived for Israel-Stewart theory, see Ref.~\cite{bd}, in the limit of vanishing $\tau_\rho$ and $\eta_\rho$. Nevertheless, we remark that simply taking $\tau_\rho=0$ is forbidden by Eq.~(\ref{trho_long}). It is possible to conclude that the inclusion of the relaxation timescale $\tau_\rho$ is essential to render the third-order theory linearly causal and stable.

\begin{figure}[ht]
\begin{center}
\includegraphics[width=\textwidth]{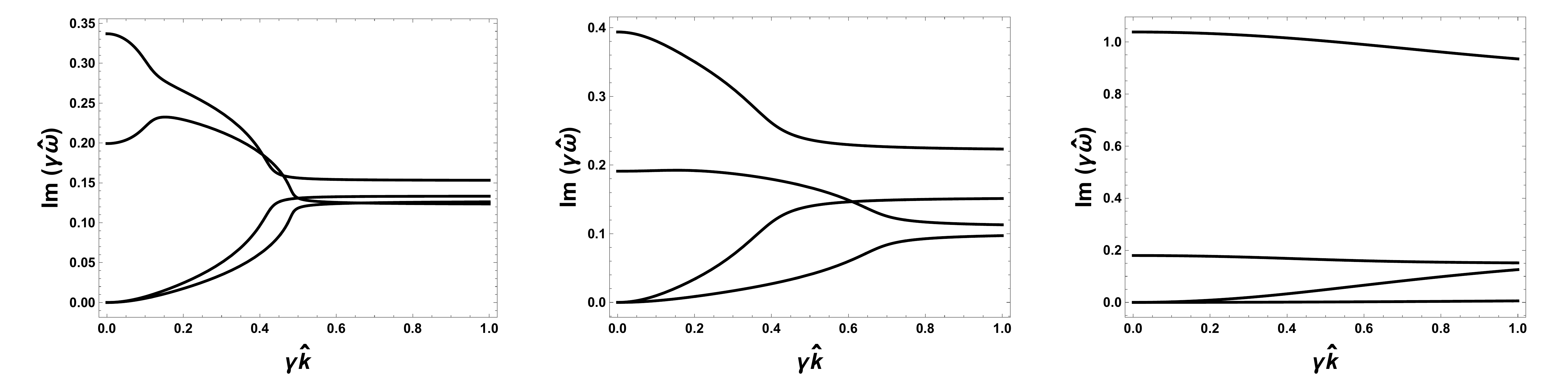}\\
\includegraphics[width=\textwidth]{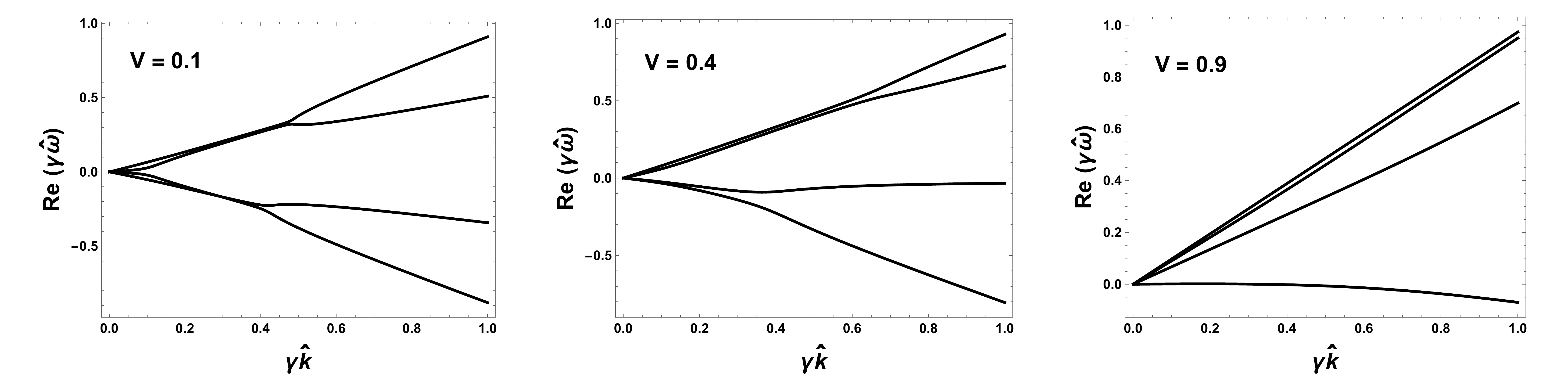}
\caption{Imaginary and real parts of the longitudinal modes, considering $\hat\eta_\rho=\hat{\tau}_\pi=5$ \cite{dnmr} and $\hat{\tau}_\rho=3$, for three different values of background velocity $V=0.1$, $V=0.4$, and $V=0.9$, in the ultrarelativistic limit $c_s^2=1/3$.}
\label{fig_relax_nonstatic_long}
\end{center}
\end{figure}

The solutions of Eq.~(\ref{eq_hyper_long_boost}) are displayed in Fig.~\ref{fig_relax_nonstatic_long}, considering $\hat\eta_\rho=\hat\tau_\pi=5$ \cite{dnmr} and $\hat\tau_\rho=3$, in the ultrarelativistic limit $c_{\mathrm{s}}^2=1/3$, for several values of background velocity. In this scenario, we note that the modes are linearly stable beyond the vanishing wave number limit. Once again, as it was also observed for the transverse modes, the imaginary part of the longitudinal modes no longer merge when considering perturbations on a moving fluid.

After carefully analyzing both transverse and longitudinal modes of the modified third-order formulation, we concluded that the inclusion of the transport coefficients $\tau_\rho$ and $\eta_\rho$ is essential to obtain a linearly causal and stable formulation. However, in order to obtain exclusively stable modes, the novel transport coefficients cannot assume arbitrary values. We then constrained the possible values they can assume in order to render causal and stable modes and noted the conditions related to the longitudinal modes of the theory are stronger than the ones obtained for the transverse modes. 

\section{Conclusions and remarks}
\label{sec_conc}

In this work, we analyzed the linear causality and stability of third-order fluid-dynamical formulations. The original formulation of third-order fluid dynamics proposed in Ref.~\cite{amaresh} was shown to be linearly unstable when perturbed around a global equilibrium state. While the modes were shown to be stable for perturbations on a static fluid, they become intrinsically unstable for perturbations on a moving fluid. The origin of this problem was mapped to the diffusionlike behavior of the modes at large values of wave number. This problem is similar to what is observed in relativistic Navier-Stokes theory, whose parabolic nature leads to acausal and unstable modes. We then observed that the instability of this theory cannot be solved by simply tuning the transport coefficients.

With the purpose of addressing this problem, we then proposed a modification of the theory developed in Ref.~\cite{amaresh} that renders it linearly causal and stable. This modified version of the third-order theory was obtained by converting gradients of shear-stress tensor, linear third-order terms that feature the equation of motion for $\pi^{\mu\nu}$, into a new dynamical variable $\nabla^{\langle\mu}\pi^{\nu\lambda\rangle}\rightarrow\rho^{\mu\nu\lambda}$ that relaxes to such gradients of the shear stress tensor. We then required that this new dissipative current satisfies a relaxation equation, introducing $\tau_\rho$ and $\eta_\rho$ as new transport coefficients corresponding to the relaxation time and effective viscosity associated with the hydrodynamic current $\rho^{\mu\nu\lambda}$, respectively. In the construction of this equation, we only accounted linear terms, since any nonlinear term would vanish and thus would not contribute to the analysis performed in this paper. 

The next step was to analyze whether the addition of the new nonconserved current $\rho^{\mu\nu\lambda}$, along with the aforementioned transport coefficients, is actually sufficient to render the theory linearly causal and stable. At this point, it is essential to perform perturbations not only on the usual hydrodynamic variables, but also in such current, $\rho^{\mu\nu\lambda}=\delta\rho^{\mu\nu\lambda}$. When calculating the dispersion relations associated with the transverse and longitudinal modes of this theory, we immediately recover the equations for the previous theory by simply taking $\tau_\rho=0$ and $\eta_\rho=\tau_\pi$. Furthermore, the modes of the modified theory do not display a diffusionlike behavior at large values of wave number and the number of modes does not increase when considering perturbations on a moving fluid, problems displayed by the original theory. We then obtained the set of conditions the transport coefficients introduced must satisfy in order for the modified theory to be linearly causal and stable. Overall, in terms of the usual hydrodynamic variables, the linear causality and stability conditions of the modified third-order theory are given by
\begin{eqnarray}
\left[3\tau_\pi\left(1-c_{\mathrm{s}}^2\right)-4\frac{\eta}{\varepsilon_0+P_0}\right]\tau_\rho&>&\frac{27}{35}\eta_\rho \tau_\pi\left(1-c_{\mathrm{s}}^2\right), \label{finaleq1}\\
3(1-\mathrm{c}_\mathrm{s}^2)\tau_\pi&\geq&\frac{4\eta}{\varepsilon_0 +P_0}. \label{finaleq2}
\end{eqnarray}

Therefore, we concluded that the transport coefficients cannot assume arbitrary values, and the novel third-order theory is linearly causal and stable as long as Eqs.~\eqref{finaleq1} and \eqref{finaleq2} are simultaneously satisfied. In particular, Eq.~\eqref{finaleq1} imposes the existence of a nonzero timescale, defined as $\tau_\rho$, as an essential requirement for both causality and stability -- in fact, these properties are related: if linear causality is satisfied, then linear stability is automatically guaranteed and vice-versa. 

A complete nonlinear third-order theory was not formally derived here and shall be investigated in future publications. In particular, it is our future goal to investigate how such novel third-order theory emerges in a microscopic derivation, taking the relativistic Boltzmann equation as the starting point. 

\section*{Acknowledgments}
The authors thank J. Noronha and S. Jeon for insightful discussions. G.~S.~D.~is funded by Conselho Nacional de Desenvolvimento Cient\'ifico e Tecnol\'ogico (CNPq) and Funda\c c\~ao Carlos Chagas Filho de Amparo \`a Pesquisa do Estado do Rio de Janeiro (FAPERJ), process No.~E-26/202.747/2018. C.~V.~B.~is funded by CNPq, process No.~140453/2021-0.

\bibliographystyle{apsrev4-1}
\bibliography{refs}

\end{document}